\documentclass[review]{elsarticle}

\usepackage{lineno,hyperref}
\usepackage{amsfonts}
\usepackage{subcaption}
\usepackage{hyperref}
\usepackage{setspace}
\usepackage[nameinlink,capitalise]{cleveref}

\begin{document}

\begin{frontmatter}

\title{Automatic Recognition of the Supraspinatus Tendinopathy from Ultrasound Images using Convolutional Neural Networks}

\author[mymainaddress]{Mostafa Jahanifar}
\author[myfifthaddress]{Neda Zamani Tajeddin}
\author[mymainaddress]{Meisam Hasani}
\author[mythirdaddress]{Babak Shekarchi}
\author[myfourthaddress]{Kamran Azema}

\address[mymainaddress]{Research \& Development Department, NRP Co., Tehran, Iran.}
\address[myfifthaddress]{Department of Biomedical Engineering, Tarbiat Modares University, Tehran, Iran.}
\address[mysecondaryaddress]{Neshat Physical Medicine and Rehabilitation Center, Karaj, Iran.}
\address[mythirdaddress]{Department of Radiology, AJA University of Medical Science, Tehran, Iran.}
\address[myfourthaddress]{Department of Physical Medicine, AJA University of Medical Science, Tehran, Iran.}

\begin{abstract}
\singlespacing
Tendon injuries like tendinopathies, full and partial thickness tears are prevalent, and the supraspinatus tendon (SST) is the most vulnerable ones in the rotator cuff. Early diagnosis of SST tendinopathies is of high importance and hard to achieve using ultrasound imaging. In this paper, an automatic tendinopathy recognition framework based on convolutional neural networks has been proposed to assist the diagnosis. 
This framework has two essential parts of tendon segmentation and classification. Tendon segmentation is done through a novel network, NASUNet, which follows an encoder-decoder architecture paradigm and utilizes a multi-scale \textit{Enlarging cell}. Moreover, a general classification pipeline has been proposed for tendinopathy recognition, which supports different base models as the feature extractor engine. Two feature maps comprising positional information of the tendon region have been introduced as the network input to make the classification network spatial-aware. To evaluate the tendinopathy recognition system, a data set consisting of 100 SST ultrasound images have been acquired, in which tendinopathy cases are double-verified by magnetic resonance imaging. In both segmentation and classification tasks, lack of training data has been compensated by incorporating knowledge transferring,  transfer learning, and data augmentation techniques. In cross-validation experiments, the proposed tendinopathy recognition model achieves 91\% accuracy, 86.67\% sensitivity, and 92.86\% specificity, showing state-of-the-art performance against other models.
\end{abstract}

\begin{keyword}
tendinopathy recognition\sep supraspinatus \sep tendon segmentation\sep tendon classification\sep positional information\sep convolutional neural networks
\end{keyword}

\end{frontmatter}


\singlespacing

\section{Introduction}

Tendons are essential parts of the human musculoskeletal (MSK) system that connect muscles to the bones to transmit the force generated by muscles. Tendons are made of collagen fibers which are packed in parallel arrays \cite{lipman2018}. 
Tendon disorders can be categorized into two class of major injuries (like full or high-grade partial thickness tears) and minor injuries (like tendinopathies) \cite{sharma2006biology,de2018ultrasound}.
Several factors can be listed as causes of tendon pathology -- intrinsic factors like malnutrition, body overweight, and aging on the one hand and extrinsic factors like accidents, performing heavy exercises/sports, applying excessive loads, repetitive microtrauma, or inappropriate workouts are the leading causes of tendon injuries \cite{sharma2006biology}. It has been reported that approximately 30\% of patients who complain about MSK pain and 30\%-50\% of sport-related injuries are diagnosed with tendinopathy \cite{lipman2018}. 

\begin{figure*}[htb]
    \centering
    \begin{subfigure}{0.31\textwidth}
        \centering
        \includegraphics[width=\linewidth]{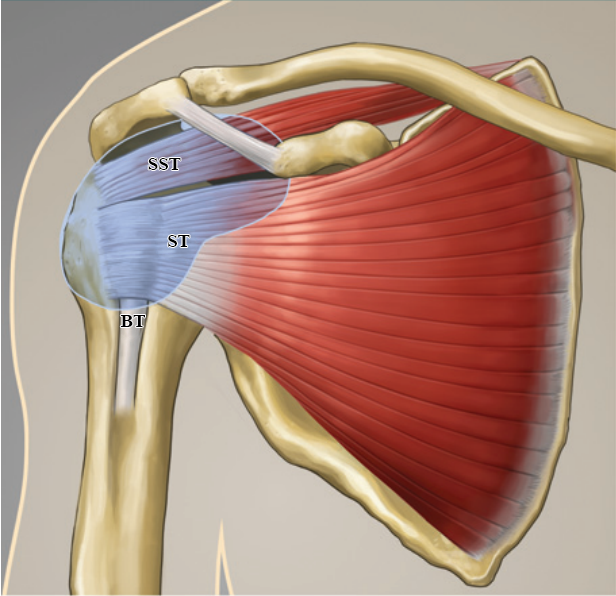}
        \caption{}
        \label{fSST1}
    \end{subfigure}%
\hfil
    \begin{subfigure}{0.68\textwidth}
        \centering
        \includegraphics[width=\linewidth]{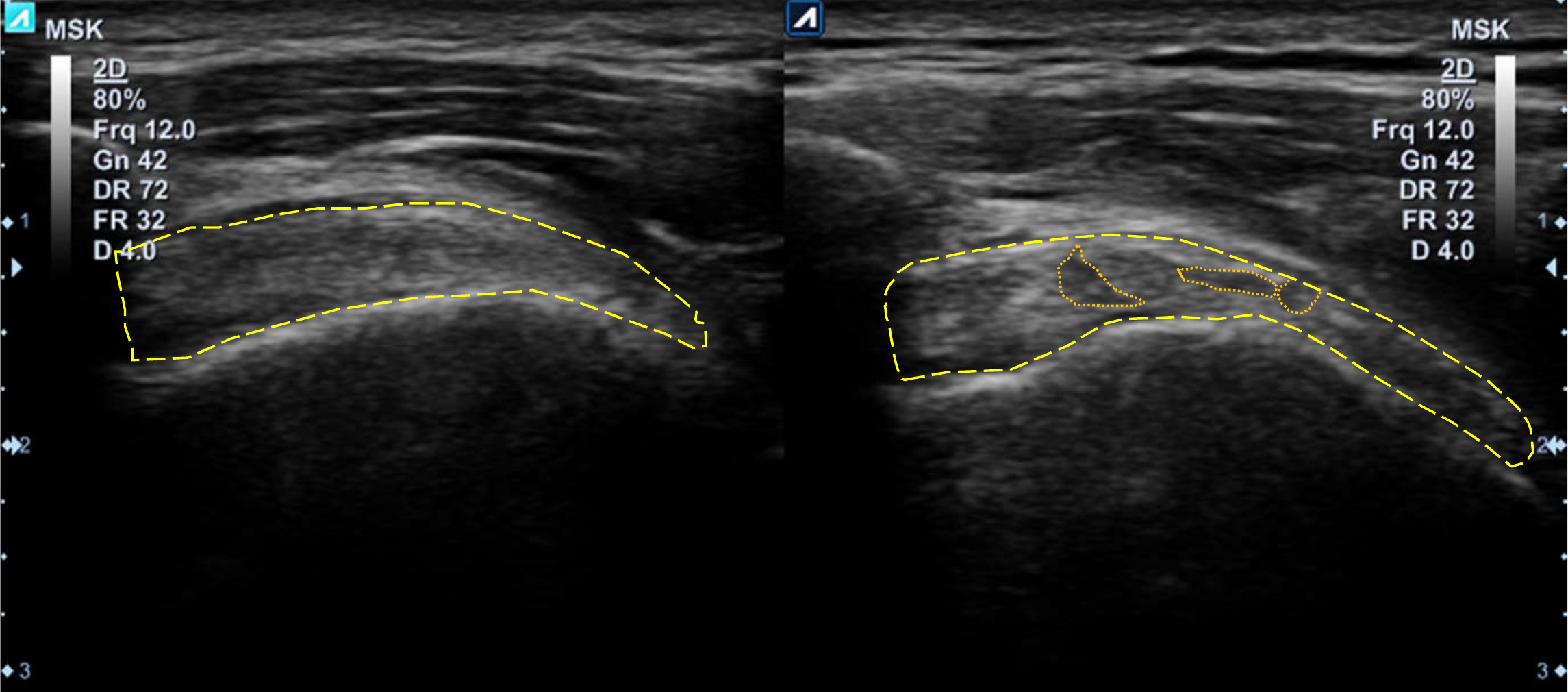}
        \caption{}
        \label{fSST2}
    \end{subfigure}
    \caption{(a) illustrates the anterior view of the shoulder anatomy, showing brachii (BT), subscapularis (ST), and supraspinatus (SST) tendons (Image courtesy of Carolyn Nowak, Ann Arbor, Mich.), and (b) depicts the modified coronal view of the SST in US images of a normal case (left) and a tendinopathy case (right). The SST area is annotated with yellow dashed lines, and orange dotted lines indicate areas of tendon pathology.}
    \label{fSST}
\end{figure*}

In this paper, automatic recognition of supraspinatus tendinopathy has been addressed, which refers to the minor disorders of the supraspinatus tendon (SST) in the rotator cuff. This tendon connects the supraspinatus muscle to the greater tuberosity of the proximal humerus and is located in the suprascapular fossa of the scapula inside the rotator cuff, as illustrated in \cref{fSST1}. Statistics have revealed that supraspinatus tendinopathy is the most commonly encountered around the shoulder joint \cite{lipman2018}. This condition causes “painful arc” syndrome and limits the patient hand/shoulder movements for several months. Early detection of supraspinatus tendinopathy can help to reduce the unfavorable pathological effects and to shorten the recovery time.

According to radiologists, the gold standard modality for tendinopathy assessment is magnetic resonance imaging (MRI) because it delivers high-resolution and high-detailed images of soft tissues. In MR images, healthy tendons reflect no signal (because they are made of collagen fibers) and they would be visible as dark areas in the tendon image. However, in tendinopathy cases, the accumulation of water or fat in the region of injured tendons causes signal reflections in MR imaging; therefore, those regions would appear brightly in the image. Depending on the severity of the injury, brighten regions may be visible as hot spots, partial textural appearance, or full tendon region glow. Based on these visual clues from the MR images, an experienced radiologist can diagnose the type of tendinopathy \cite{chang2009imaging}.

However, considering the common situations that patients are afflicted with tendon injuries and costs of MRI, it is always preferred to use a real-time, mobile, and cost-efficient modality like ultrasound (US) imaging \cite{stevic2013us}. Furthermore, recent studies confirmed that US imaging could be considered as a trustworthy replacement for MRI in the detection of tendinopathies \cite{de2018ultrasound, stevic2013us, saraya2016, scott2018diagnostic}. Nonetheless, US imaging modality has its drawbacks. For example, US images show less resolution and details for soft tissues while carrying more noise, i.e., having a lower signal to noise ratio (SNR) in comparison with MR images \cite{faust2018comparative}.
US images of  SST for both left and right rotator cuffs of a person are depicted in \cref{fSST2}. The left image of \cref{fSST2} corresponds to a normal SST, while the right image shows a tendinopathy case. Unlike MRI, injured regions of the tendon would appear darker than the healthy tissue in US imaging due to water/fat accumulation  \cite{stevic2013us}. Also, the geometry of the abnormal case is changed especially in the peripheral region where tendon boundary is not smooth anymore, and it has become thinner in comparison to the normal case \cite{stevic2013us}. However, textural changes caused by tendinopathy in US images are not as apparent as they are in  MR images. As \cref{fSST2} illustrates, texture differences between the normal and tendinopathy cases are very subtle and hard to detect (visual markers related to tendinopathy are annotated with orange dotted lines in \cref{fSST2}).

Vague textural patterns alongside image deficiencies that come with US modality make the image interpretation and diagnosis harder for physiatrists and radiologists \cite{matthews2018classification}. Therefore, these hurdles increase demands for a reliable computer-assisted diagnostic (CAD) system, which can help medical specialists in tendinopathy recognition.

CAD-based automatic recognition methods in the field of medical images usually consist of three main steps \cite{faust2018comparative,tajeddin2018melanoma}: segmentation of the desired object, extracting features from the segmented region and identifying the type of the object by classifying the extracted features through different machine learning tools. Tendinopathy markers US images are far more complicated to be formulated deterministically or to be detected using low-level image features \cite{matthews2018classification}. Ergo, we decided to use sophisticated deep neural networks to approach this problem. In recent years with the advent of deep neural networks, especially convolutional neural networks (CNN), the need for manual feature engineering has been fading out. CNNs are potent tools that if provided with enough training data and configured correctly, can automatically extract local and abstract features from the input image in a multi-level framework \cite{lecun2015deep}.

Unfortunately, automatic tendinopathy recognition has not been widely discussed in the medical image analysis community, and the main reason seems to be the absence of a publicly available data-set for the task. At the time of writing this paper, we could find only one research devoted to supraspinatus tendon segmentation by Gupta et al. \cite{gupta2014curvelet} and another one dedicated to automatic supraspinatus tears detection \cite{chang2019quantitative}. Gupta et al. \cite{gupta2014curvelet} proposed to apply a curvelet transformation to extract basic features followed by energy analysis of the extracted features to build an initial mask of the tendon. The initial mask is then refined by performing morphological operations. The final segmentation map is then created by fitting a polynomial curve on the refined mask. Their method was able only to extract a rough region comprising supraspinatus tendon, but not the actual boundaries of it. 
Chang et al. \cite{chang2019quantitative}  proposed to use hand-crafted features in a binary logistic regression classifier to identify tendons with tears. The incorporated features in their research were intensity and tecture descriptors. In \cite{chang2019quantitative}, intensity features were the first to fourth central moment of the histogram generated from tendon region pixels. On the other hand, textural features are extracted from the gray-scale co-occurrence matrices (GLCM) in different directions \cite{chang2019quantitative}. Chang et al. \cite{chang2019quantitative} showed that the combination of intensity and textural features results to the best recognition performance. However, in \cite{chang2019quantitative} authors extracted features from the tendon/ or tear areas which were previously annotated by an expert. Therefore, it can be said that their research is limited to the analysis of the performance of intensity and textural features for classifying annotated regions of supraspinatus tendon into tear or tendinopathy groups.

Other methods in the literature have done research on other tendons rather than SST. In a similar research to  \cite{chang2019quantitative}, Meiburger et al. \cite{meiburger2018quantitative} assessed the performance of 90 features, extracted from manually segmented regions, in a multivariate linear regression model for the patellar tendon abnormality detection. Their study showed that hand-crafted features can perform reasonably for tendon abnormality detection from US images. However, the dataset that they used for model training and evaluation consisted of only 14 cases, which indicates that results may be unreliable or very prone to  over-fitting. 

Chuang et al. \cite{chuang2014model} introduced a model-based method for tendon segmentation from the US image of fingers. Their segmentation model was optimizing an energy function comprised of shape and texture features. Chuang et al. \cite{chuang2014model} borrowed a method similar to famous active appearance model (AAM) \cite{ cootes2001active} for their model optimization and tendon segmentation. Recently, an automatic Achilles tendon (AT) abnormality detection method has been proposed by Benrabha et al. \cite{benrabha2017}. Classically, they enhanced the image quality through pre-processing steps and then divided the image into several blocks (ROIs) to transform the image classification task into an ROI classification task. In other words, they skipped the segmentation task and tried to decide about the image class (normal/abnormal) by analyzing its ROIs. Various features representing the texture and intensity of each ROI were extracted to be classified using different classifiers. They reported that an ensemble classifier of bagged trees was able to outperform other classifiers in AT classification task. Because their methodology greatly depends on the manual AT region cropping in the pre-processing step, this algorithm is inappropriate for automatic tendinopathy detection.

Unlike the limited literature on tendinopathy recognition, there is a vast literature around CNN-based methods for segmentation and classification tasks. Most of them were applied to the medical images and achieved promising results \cite{litjens2017survey}.

In the current research,  a data set of US images have been acquired from SST comprising both normal and tendinopathy cases in a reliable and repeatable framework which encompass double-modality validation for tendinopathy cases. A novel methodology to recognize tendinopathy from healthy cases has been proposed. Our method comprises of two novel CNN architectures for tendon segmentation and classification tasks. The segmentation network is based on transfer learning concepts and uses multi-scale convolutional blocks. It is always a good practice to look for diagnostic markers in positions that they usually occur; that is what most radiologists do when they are assessing a new field of view.
Fortunately, unlike usual convolutional neural networks, our spatial-aware deep learning framework (which is constructed by introducing tendon positional information maps to the network) can mimic this ability and recognize where a specific textural pattern occurs, hence, resulting to a more precise detection of tendinopathy.

\section{Data set acquisition and description}
\label{sec:data-set}
In this research,  a supervised learning framework has been used for automatic recognition of tendon pathology. To this end, we need an annotated data set comprising both normal and tendinopathy cases. Therefore, a data set of 100 US images taken from supraspinatus tendon has been acquired. 

Our data acquisition routine enforces that all tendinopathy cases should be validated by magnetic resonance (MR) imaging. Accordingly, the routine for data acquisition from normal cases (cases that do not show any clinical symptoms or have no pain in their shoulder area) involves only US imaging, whilst for cases that express pain or usual tendinopathy symptoms, first MR imaging of the shoulder area is prescribed for the patient which will be analyzed by an experienced radiologist. If radiologist observed and approved any sign of tendon pathology based on the MR image, that case would be considered as positive tendinopathy, and US imaging is carried out on it with the same protocol.

US image capturing was done based on a reliable and repeatable protocol \cite{jacobson2011shoulder}. Two different centers have been involved in data acquisition, and each utilized US imaging devices from different vendors, E-CUBE 7 from the ALPINION and SonoAce X8 from Medison. Both instruments provide the same resolution and are equipped with a 10 MHz linear transducer probe for musculoskeletal imaging purposes. During the imaging, the transducer placement with the patient’s shoulder was done in a modified Cross position \cite{jacobson2011shoulder}, in which the patient’s hand must be placed on the buttock or hip region. By approaching this patient positioning, radiologists can capture a long-axis view of the supraspinatus tendon in the coronal plane. This view is very important for supraspinatus imaging because not only anatomic surfaces of the tendon can be identified, but also this view makes probable tendon inflammations or tears visible \cite{jacobson2011shoulder}. 

Acquired data set for this research can be divided into 70 healthy and 30 tendinopathy cases. In the whole data set, the male/female ratio of cases’ gender is about 7/3, more explicitly, 22 cases out of 70 normal cases are females and remaining are males. About the same statistics hold for tendinopathy cases, i.e., 12 out of 30 tendinopathy cases are females. In terms of different device statistics, 57 cases out of 100  were captured using E-CUBE 7 device and other 43 cases using  “SonoAce X8”. The first group has about 21\% positive tendinopathy, i.e., 12/45 portion in the positive/negative population. For the second group, positive-negative data partitioning is as follow: 18 cases are tendinopathy while 25 are normal showing about 42\% positive cases in the population.

Please note that the above-reported class populations are achieved after pruning certain individuals from the trial. In other words, during the data acquisition, the prevalence of tendon pathology in the painful group was not 100\%; therefore, we employed a case rejection policy explained as following. For cases that had painful symptoms, both MR and US should be carried out, and the radiologist would decide if those cases have any tendinopathy in the supraspinatus tendon or not. If not, those cases would be rejected from the trial altogether. In other words, a tendinopathy rejected case would not be considered as a normal case. This has been done to account for errors in imaging procedure and radiologist decision. On the other hand, for cases without painful symptoms, if radiologist notices any sign of abnormality in the US image, that case should be validated by MR imaging. If MR image shows tendinopathy symptoms in the supraspinatus tendon, that case would be categorized in the tendinopathy class other-wise it would be eliminated from the trial. Based on this rejection policy, from 75 persons of the painless group that participated in the trial, only 5 showed tendinopathy signs in the US image, and all were rejected to be categorized in either normal or abnormal groups. On the other side, from a group of 43 patient with painful symptoms, only 30 get verified to have supraspinatus tendinopathy, and the remaining 13 cases were rejected. Because, source of the pain for rejected cases was not supraspinatus tendinopathy, i.e., either supraspinatus tendon tears were present, or other tendons in the rotator cuff were injured.

In order to use the acquired data in a supervised learning framework, images must be labeled to serve as training and evaluation references. Thus, the boundaries of the SST in the image are manually and carefully annotated by an expert radiologist to form references for the SST segmentation task. For the classification (recognition)  task, the reference was a joint decision of two radiologists and a physiatrist (who have over 15 years of experience) based on both US and MR images. Normal (healthy cases) were decided only by using US image information, while tendinopathy cases were verified by MR images, i.e., both US and MR images were incorporated in the decision making.

\section{Methodology}
\label{sec:methods}
Based on the nature of CNNs that extract local features using trainable convolutional kernels, an important property of them is image classification without the need for segmenting particular objects. However, there are researches in medical fields that show segmenting the desired object in the image before image classification can extremely improve the recognition performance \cite{jahanifar2019supervised}. Therefore, we first segment the tendon from the US image, transform the segmentation map into positional information maps, and use those maps in our recognition framework. 

 A novel CNN model has been proposed for the segmentation task, which benefits from exclusive structural blocks. For the classification task, a CNN-based classification framework has been introduced that can operate with different base models. This framework is equipped with positional information of tendon in the input to boost its performance. However, the primary challenge in this research is the scarcity of data. Usually, CNNs need lots of data (often more than 1000 samples per class) to reach a good convergence and generalization during the training phase \cite{lecun2015deep}, but access to such a big data set of SST US images is costly and time-consuming. Thus, we embed modern techniques in our methodology to tackle the data shortage problem.  Details around the proposed methods for tendon segmentation and recognition tasks are fully described in the following subsections.

\subsection{Dealing with data shortage problem}
	
In training deep learning models, lack of training data may lead to over-fitting \cite{lecun2015deep}. However, acquiring an extensive data set in medical cases is not readily achievable, similar to the situation where abnormal cases are validated with a second expensive imaging modality. Therefore, we propose to use knowledge transferring, transfer learning, and data augmentation techniques to tackle the problem of data shortage and making our networks more generalizable and robust against over-fitting.

\subsubsection{Knowledge transferring}
Zen et al. \cite{nasnet} proposed a network architecture search (NAS) engine to obtain optimum building blocks automatically. They proposed to perform architecture search on a simple data set. After obtaining the optimal building blocks, they used them to design more advanced network architectures (NASNet) suitable for working with more complex data sets. This approach is called knowledge transferring, where knowledge of network design obtained from more straightforward problems is used to solve harder ones.

\begin{figure*}[h]
	\centering
    \includegraphics[width=1\textwidth]{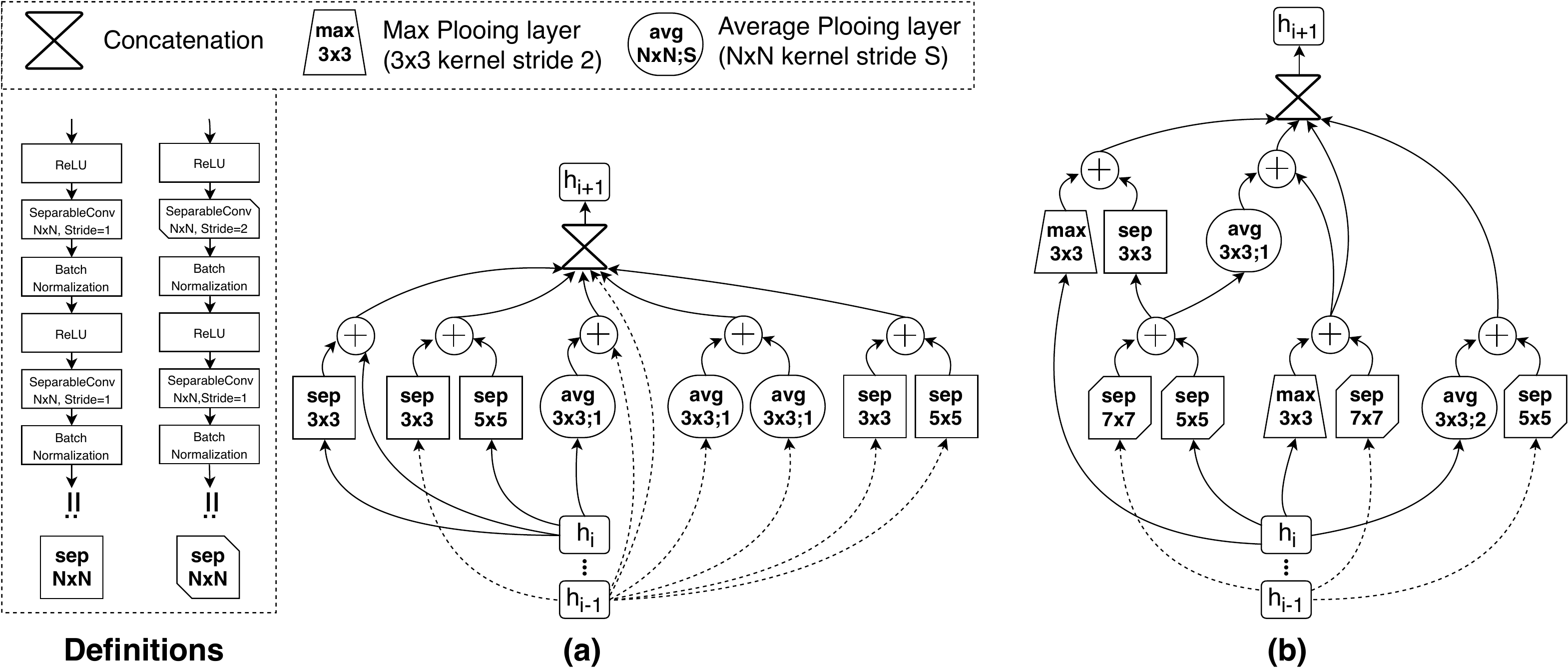}
	\caption{Structuring cells for realising knowledge transferring adopted from \cite{nasnet}: (a) the Normal cell, (b) the Reducing cells}
\label{fNormReducBlock}
\end{figure*}

Our method benefits from the knowledge transferring concept as well. Essential building blocks from NASNet \cite{nasnet}, the “Normal” and the “Reducing” cells, are borrowed to construct our tendon segmentation and tendinopathy recognition networks. A Normal block, as illustrated in \cref{fNormReducBlock}a applies separable convolutions \cite{xception}, and average pooling \cite{lecun2015deep} kernels with different sizes on its input and then combines their results in order to construct the block’s output, which is the concatenation of the feature maps resulted from different operations. The Reducing cell depicted in \cref{fNormReducBlock}b incorporates the stridden version of separable convolution, average pooling, and max-pooling layers. If stride value in each of those layers set to 2, their output size would become half of their input size in both horizontal and vertical dimensions.  The Reducing cell can extract features from the input and resize (down-sample) output feature maps in half at the same time. The way that different layers are merged to build up the Normal and Reducing cells are depicted in \cref{fNormReducBlock}.

\subsubsection{Transfer learning}
Transfer learning is when kernel weights of a network trained on a specific data set are used to work on another data set coming from different distribution \cite{van2015transfer}. By initializing the network kernels based on the pre-trained weights instead of initializing the weights randomly, the training procedure of the network on the desired data set starts from a more reasonable point, usually leading to faster and more generalizable convergence \cite{van2015transfer, jahanifar2018segmentation}.

In our proposed transfer learning framework, the base model for feature extraction is NASNet-A (6 @ 4032) \cite{nasnet}, which is a powerful model outperformed other state-of-the-art methods in image classification on ImageNet data set \cite{deng2009imagenet}. The NASNet base model that we have incorporated in our segmentation or recognition networks is initialized with the pre-trained weights of the ImageNet data set. This data set has over 1 million images of natural scenes \cite{deng2009imagenet}, which can guarantee to achieve well-generalized pre-trained kernel weights. However, compared to natural images, US images are different in appearance and have some intrinsic characteristics and deficiencies that make their inference more challenging. A two-phase training approach based on transfer learning has been proposed to adapt our network to the new domain (US images).

In the first phase, before training the network with our data set, we propose to train the network on a relatively bigger data-set of US images, the “Ultrasound Nerve Segmentation Challenge” (UNS) data set \cite{uns}. The UNS data set consists of 5635 US images from the human neck area accompanied by ground truth segmentation of the Brachial Plexus’ nerve structures in them \cite{uns}. Training our networks on this relatively big data set help their convolutional kernels recognizing US image structures and appearance better. 
The second phase of training is to fine-tune the network weights on available SST data set. Because our network has been trained on extensive data set of UNS beforehand, now it can easily capture the essence of the tendon US images and extract relevant features to detect the tendon area in the image correctly. 

\subsubsection{Data augmentation}
Apart from the knowledge transferring and transfer learning, using data augmentation techniques (DAT) has been considered to increase the number of training images synthetically. Using DATs would make the network robust against deficiencies and variations that may appear in the unseen test data \cite{jahanifar2018segmentation}. During the training of the segmentation network,  these augmentation techniques are utilized on-fly (extents of each augmentation are reported in parentheses): intensity scaling (multiplying the image intensities by a random scale in the range 0.7 to 1.3), contrast adjustment (randomly adjusting the contrast using linear intensity transformation by extent of 20 gray levels), illumination gradient adjustment, vertical and horizontal flipping, rotation (by a random extent in the range of -20 to 20 degrees), zooming (geometrically scaling the image by a random extent in the range of 0.8 to 1.2 of the original image size), translation (up to 0.1 image width or height), shearing (up to 0.15 image width or height), elastic deformation, and sharpness adjustment (using unsharp masking or Gaussian smoothing methods), as described in \cite{jahanifar2018segmentation}.

However, data augmentation is more critical and sensitive in the recognition task, which means that we are not allowed to apply augmentation techniques that change the nature or type of a tendon. Hence, for the recognition task, only these augmentation techniques have been implemented: vertical flipping, rotation, zooming, translation, shearing, and adding noise.

\subsection{Tendon segmentation}
Identifying the SST region from US images is an arduous task due to complex texture, low SNR, and high variation in tendon appearance. Therefore, achieving tendon segmentation through classical (unsupervised) algorithms like thresholding is not possible. Since US images are gray-scale images and SST does not show any saliency in the image, supervised saliency detection algorithms like \cite{jahanifar2019supervised} are not practical either. Hence, we utilize CNNs which can extract multi-scale important features and create the segmentation map in an end-to-end framework. In the following sections,  our proposed segmentation network, NASUNet, will be explained.

Beside our segmentation network, in this research, five other well-known segmentation models have been tested on the SST data set. Four of these models are FCN-8 \cite{fcn}, U-Net \cite{unet}, SegNet \cite{segnet}, and DeepLab v3 \cite{deeplab}, which all are based on CNNs and share a general encoder-decoder paradigm in their architecture. The FCN-8 \cite{fcn} was the first end-to-end CNN model introduced for semantic segmentation, which directly enlarges the final feature maps of a classification network. Ronneberger et al. \cite{unet} proposed a network specially tailored for medical image segmentation. In U-Net \cite{unet}, both encoding and decoding paths use convolutional layers and are designed in a multi-level manner. Skip connections are also incorporated to elevate the performance for high-resolution predictions. So far, U-Net and its derivatives have been one of the most success full networks in different applications of medical image segmentation field \cite{litjens2017survey}. For example, SegNet \cite{segnet} is one of the U-Net variants with minor changes which was able to achieve impressive results on semantic segmentation tasks. In more recent research, Chen et al. \cite{deeplab} proposed DeepLab v3, which uses atrous convolutions with different scales on the last feature maps of the encoder  (which is an Xception model \cite{xception}). Use of multi-scale convolutions helped DeepLab model achieve state-of-the-art results on semantic segmentation task. Applying DeepLab v3 \cite{deeplab} model on medical image segmentation task also showed promising results \cite{jahanifar2018segmentation}. 

\subsubsection{Network architecture}
Our proposed segmentation architecture is based on an encoding-decoding design. In which, basic blocks in the encoding path extract features in a multi-level paradigm, lower levels contain high-resolution information, but higher levels represent more abstract feature maps, as their resolution decreases. On the other hand, feature maps’ resolution increases in the decoding path by enlarging the extracted feature in the same number of levels. 

\begin{figure*}[htb]
	\centering
    \includegraphics[width=1\textwidth]{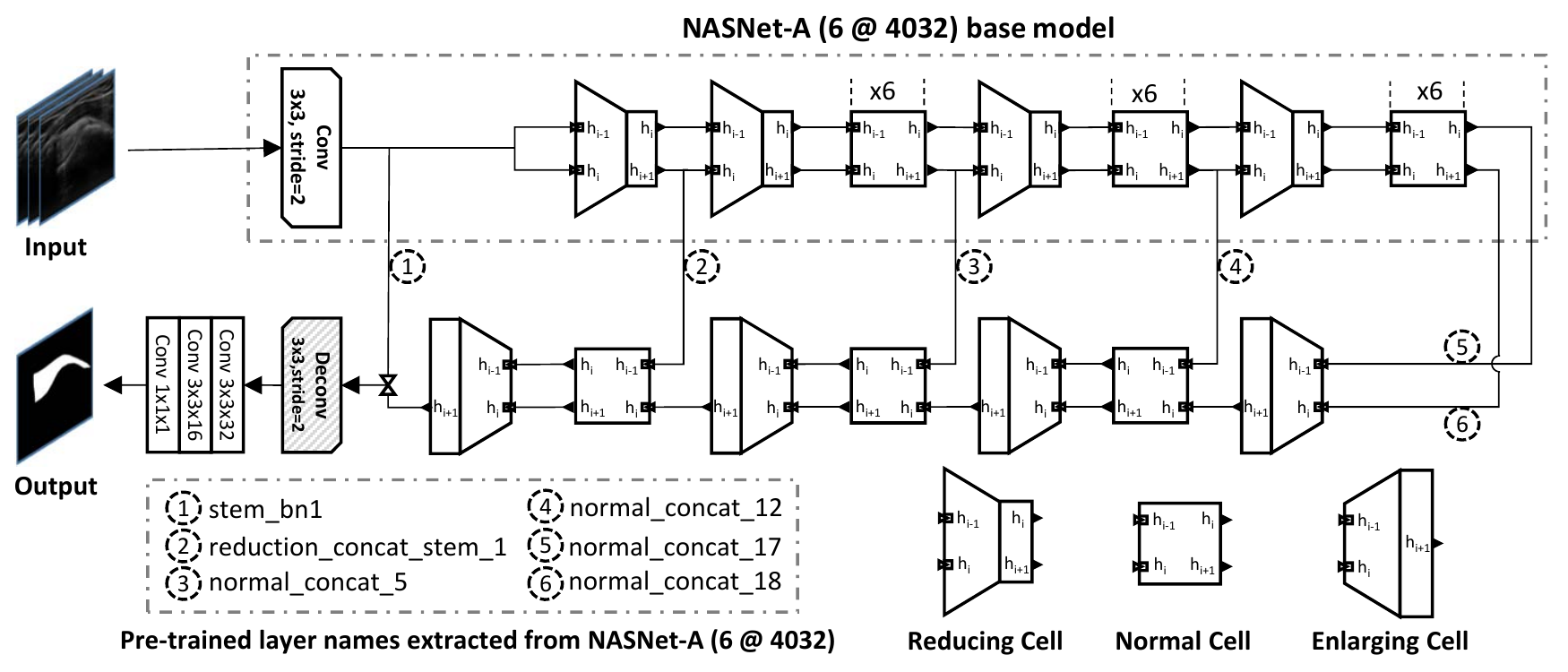}
	\caption{Network architecture of the proposed segmentation model, NASUNet. Enconding path of the NASUNet is equal to NASNet-A (6@4032) model \cite{nasnet} which comprises Normal and Reducing cells (introduced in \cref{fNormReducBlock}) and the decoder part  incorporates the novel  Enlarging cell (introduced in \cref{fEnlgBlock}).}
\label{fSegArch}
\end{figure*}

As depicted in \cref{fSegArch}, in our segmentation architecture, three main building blocks are utilized: Normal, Reducing, and Enlarging cells.  The Normal and Reducing blocks are directly borrowed from NASNet \cite{nasnet}. Inspired by U-Net and NASNet models, we will call our proposed segmentation network "NASUNet" from the hereafter.

\begin{figure*}[h]
	\centering
    \includegraphics[width=0.6\textwidth]{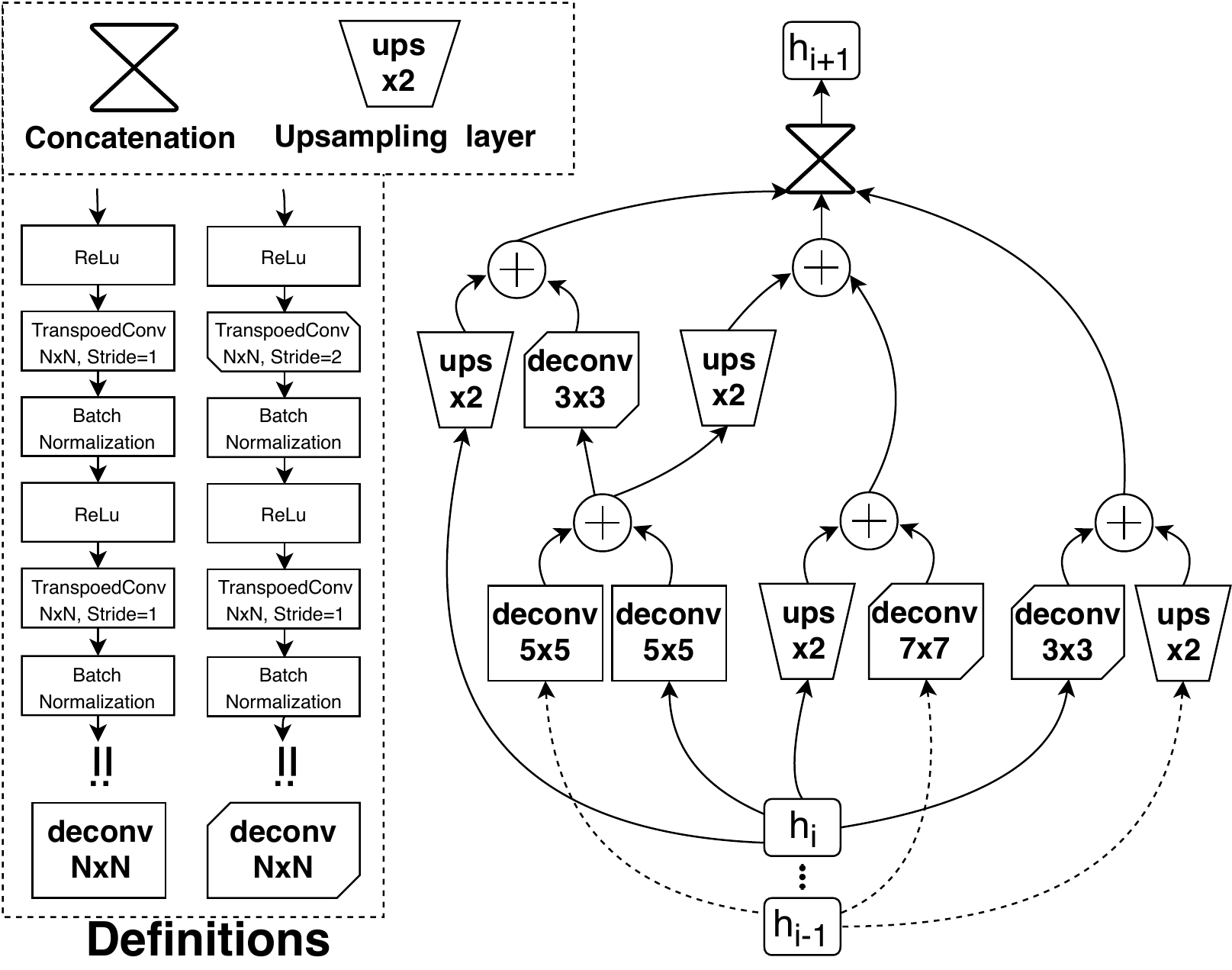}
	\caption{Architecture of proposed Enlarging cell. This structuring block is used in the decoding path of the segmentation network (\cref{fSegArch}) to enlarge the abstracted feature maps, increasing their resolution while enriching their context.}
\label{fEnlgBlock}
\end{figure*}

\paragraph{Enlarging cell} Unlike the original U-Net \cite{unet} or other segmentation networks \cite{segnet,deeplab} that use only stridden deconvolution or interpolation layers to upsample the feature map sizes, a more advanced building block has been proposed that incorporates various upsampling kernels. This building block is named as "Enlarging cell" which utilizes all of the stridden deconvolution (transposed convolution) \cite{dumoulin2016guide}, interpolation (nearest neighbor method), and average pooling layers \cite{lecun2015deep} simultaneously. Unlike stridden convolution (with a stride of 2), stridden deconvolution (with a stride of 2) increases feature map sizes \cite{dumoulin2016guide}. Configurations and connections between different constituent layers of an Enlargement cell are depicted in \cref{fEnlgBlock}. An important characteristic of the novel Enlarging cell is that it utilizes kernels with different sizes, enabling it to capture multi-scale information while up-sampling feature maps.

From \cref{fNormReducBlock,fEnlgBlock}, it is obvious that apart from separable convolution, deconvolution, average pooling, and max-pooling layers, our proposed building blocks consist of three types data streams: $h_{i-1}$, $h_{i}$, and $h_{i+1}$. In a network architecture, the $h_{i+1}$ is the output stream from the current building block, but $h_{i}$ and $h_{i-1}$ (which serve as block inputs) are outputs streams from the previous block in the network. Normal and Reducing cells accept  $h_{i}$ and  $h_{i-1}$ in their input and pass the  $h_{i}$ and $h_{i+1}$  in their output ($h_{i}$  stream remains untouched through the block to appear in the output). 

The concept of stream connections is clearly shown in the proposed network architecture of \cref{fSegArch}. In the encoding path, $h_{i}$  and $h_{i+1}$ outputs of an intermediate block would serve as $h_{i-1}$ and $h_{i}$  inputs for the next block, respectively. Connections are different for the Normal cell in the decoding path of the proposed segmentation network. Decoding path benefits from our novel Enlarging cell to up-sample feature maps sizes in a context-preserving approach. An Enlarging cell accepts both outputs from its previous block as inputs but gives only $h_{i+1}$ stream in output. For the next normal block right after an Enlargement cell, the  $h_{i}$ input is inherited from the previous Enlarging cell, but its $h_{i-1}$  input is taken from the  $h_{i+1}$ output of a Normal cell in the same relative level of the encoding path (see \cref{fSegArch}). Doing so, we can realize skip-connections, which will pass the missing high-resolution information from encoding path to the decoding path and also facilitating the gradient flow through the network, which will help to avoid the gradient vanishing effect during the training phase \cite{unet,lecun2015deep}.

It is crucial to note that in our proposed segmentation network, the architecture of the encoding path is entirely equal with NASNet-A (6 @ 4032) model \cite{nasnet}, albeit, without its top layers, i.e., average pooling and dense layers. Therefore, we can benefit from transfer learning.  Transfer learning reduces the training time (number of epochs) and guarantees faster and more reliable network convergence \cite{van2015transfer}.

At the beginning of the encoding path or NASNet-A, a 2-stridden convolution layer is placed that reduces the size of the input image. Therefore, at the end of the decoding path,  a 2-stridden deconvolution layer has been implemented to upsample the feature maps and reversing that effect. The final segmentation map is created by concluding 3x3 and 1x1 convolutional layers. Unlike all other convolutional layers in the network that are accompanied with a ReLU activation layer, a sigmoid activation function is applied on the output of the final 1x1 convolutional layer in order to make pixel values of the segmentation map bounded to 0 and 1 interval.


\subsubsection{Loss function}
Tendon segmentation can be considered as a pixel classification task, which is a dense prediction problem. In this case, the number of positive pixels (pixels that belong to the tendon area) and negative pixels (zero values pixels in the ground truth which belong to the background region) would be greatly unbalanced. Therefore, loss functions like binary cross-entropy are not suitable for this application \cite{jahanifar2018segmentation}. We propose to use a hybrid loss function, which tries to maximize the similarity between the network predictions and ground truths. Our proposed segmentation loss is defined as below:
\begin{equation}
    {\mathcal{L} = 1 - \frac{{2\sum\limits_{i,j} {{p_{ij}}{g_{ij}}}  + \varepsilon }}{{\sum\limits_{i,j} {p_{ij}^2}  + \sum\limits_{i,j} {g_{ij}^2}  + \varepsilon }} - \frac{1}{N}\sum\limits_{i,j} {{g_{ij}}\log ({p_{ij}})}}
\label{eq:Loss}
\end{equation}

In which ${{p_{ij}}}$ is the value of every (i,j) pixel in the prediction map, ${{g_{ij}}}$ is pixel values from the ground truth mask, $\varepsilon$ is small constant for avoiding division by zero which is set to $\varepsilon=1$, and $N$ is the number of pixels in the segmentation map.

As \cref{eq:Loss} shows, the proposed loss function has consisted of two parts: a Dice-like similarity index and a binary cross entropy (BCE) function which is averaged on all pixels of a single prediction. The Dice part of loss function makes the training process converge faster and more stable. It is also robust against the pixels imbalanced class population. Dice loss is appropriate to learn general parts of the tendon in the segmentation output whereas BCE penalizes the loss for pixel-wise errors. Incorporating BCE criterion alongside using squared values of ${p_{ij}^2}$ and ${g_{ij}^2}$ in the denominator of Dice part of the loss function, instead of their absolute values, would lead to a more precise boundary estimation and more soft predictions on uncertain areas. The proposed loss function is fully differentiable to be integrated into a gradient descent algorithm  within  the back propagation: 

\begin{equation}
{\frac{{\partial L}}{{\partial {p_{ij}}}} =  - \frac{{(2\sum\limits_{i,j} {{g_{ij}}} )(\sum\limits_{i,j} {p_{ij}^2}  + \sum\limits_{i,j} {g_{ij}^2} ) - (2\sum\limits_{i,j} {{p_{ij}}} )(2\sum\limits_{i,j} {{p_{ij}}{g_{ij}}}  + \varepsilon )}}{{{{(\sum\limits_{i,j} {p_{ij}^2}  + \sum\limits_{i,j} {g_{ij}^2}  + \varepsilon )}^2}}} - \frac{1}{N}\sum\limits_{i,j} {\frac{{{g_{ij}}}}{{{p_{ij}}}}}}
\label{eq:Loss_Diff}
\end{equation}

\subsubsection{Training procedure}
The training procedure is done with an Adam optimizer \cite{kingma2014adam}. For training on UNS data set the learning rate for the encoding path was set to 0.001, while for the other randomly initialized layers in the decoding path it was set to 0.003. This first phase of training longed for 200 epochs. In the second phase of training, for training on tendon US data set, the learning rate of 4e-4 has been employed for all layers and training continued for 100 epochs. 

In order to reasonably validate our segmentation network,  5-fold cross-validation framework has been used. In this framework, the whole data set is divided into five folds (groups), and the training procedures are repeated five times. In each time, images from four folds are used to train a model, and that model is then tested on the images from the left-out fold. By doing that, all images in the data set will serve once as the test image. In other words, at the end of the cross-validation experiment, network predictions for each image in the data set would be present. Our models have been evaluated in a 5-fold cross-validation framework to reduce the risk of over-fitting and generating a basis for optimal hyper-parameters estimation.

In all segmentation experiments, the input to the network is an image with 3 channels that each channel is the replication of the US image (because pre-trained NASNet expects three channels as the input). Then, it is pre-processed before feeding to the network like the authors did in \cite{nasnet} i.e., standardizing the input image data set to have zero mean and unit standard deviation. 

\subsubsection{Post-processing}
We kept the post-processing operations as simple as possible. First, a binary mask of the network prediction is created by applying threshold $T=0.4$ on the prediction map, and then the biggest object in the image is kept by size filtering in order to create the final segmentation output. 
The optimal threshold value, $T$, is selected by changing the candidate threshold value in a specific range of values and evaluating the resulted binary mask. 

\subsection{tendinopathy recognition}
In this section, our proposed model for tendon type classification will be explained. As mentioned before, CNN based methods prove to be very powerful for classification applications. We establish our recognition method based on CNNs. First, a novel manner for integrating positional information into the recognition system is explained, then our proposed network architecture for the tendinopathy recognition will be explained, and finally,  details around the training procedure and other techniques that have been incorporated in the proposed framework will be expanded.

\subsubsection{Incorporating positional information}
\label{sec:posInf}
The relative location of different structures in the object of interest is an important factor for the recognition task. For example, knowing that tendon inflammations usually happen in the end stretches of a tendon can help to achieve a better diagnose by paying more attention at those positions. Conventional CNNs are unable to capture the relative position of important structures in the object; in fact, they can only detect the presence of a specific structure in the image \cite{lecun2015deep}. Recent spatial-aware models, on the other hand, can show better performance for different tasks and various applications \cite{alemi2019spanet}.

For our application, the network awareness of the relative positioning of essential elements in the object is vital. Therefore, two feature maps are engineered that when coupled together, will enable the network addresses the position of every point inside the object of interest. The proposed positioning scheme is based on the polar system, in which every position inside the object is determined via two parameters: distance (radius) from the object origin ($\bf{r}$) and angle of deviation from the object orientation (${\bf{\theta}}$). These two parameter maps are built based on the resulted segmentation map.

Having the tendon mask, ${\bf{M}}$, that specifies the tendon region in the image, $\Omega$, as a set of pixel locations that are equal to 1, i.e.:
\begin{equation}
    {\Omega  = \{ (x,y) \in {\mathbb{R}^2}\left| {{\bf{M}}(x,y) = 1} \right.\} },
\end{equation}
the center of the tendon region is considered as the origin of our polar positioning system, ${{\bf{O}}_{x,y}}$:
\begin{equation}
    {{O_x} = \frac{1}{m}\sum\limits_{{x_i} \in \Omega } {{x_i}\quad ;\quad } {O_y} = \frac{1}{m}\sum\limits_{{y_i} \in \Omega } {{y_i}\quad } },
\end{equation}
where $m$ is the number of positive pixels in the mask region (number of valid members in the $\Omega$ set) and $({{x_i},{y_i}})$ are their horizontal and vertical positions, respectively. Parameters of the polar positioning system $(\bf{r},\Theta)$ are defined for every point (pixel) inside the tendon area, $({x_i},{y_i})$ in $\Omega$, as follow:

\begin{equation}
{{\Theta} = \{ {\theta _{xy}}|{\theta _{xy}} = \arctan (\frac{{y - {O_y}}}{{x - {O_x}}}) - \alpha \;\forall \;(x,y) \in \Omega \} },
\label{eq:pos1}
\end{equation}
\begin{equation}
    {{\bf{r}} = \{ {r_{xy}}|{r_{xy}} = \sqrt {{{(x - {O_x})}^2} + {{(y - {O_y})}^2}} \;\forall \;(x,y) \in \Omega \} }.
\label{eq:pos2}
\end{equation}

In \cref{eq:pos1}, $\alpha$ is equal to tendon region orientation, which is the angle between the major axis of the ellipse that has the same second-moments as the tendon object in the image and the horizontal axis. These two positioning maps, $(\bf{r},\Theta)$, have all information needed to specify the position of every pixel in the image. However, values of these two maps for every pixel position outside the tendon mask region are set to -1, in order to distinguish the object border in both positional maps.

\begin{figure*}[!h]
	\centering
    \includegraphics[width=1\textwidth]{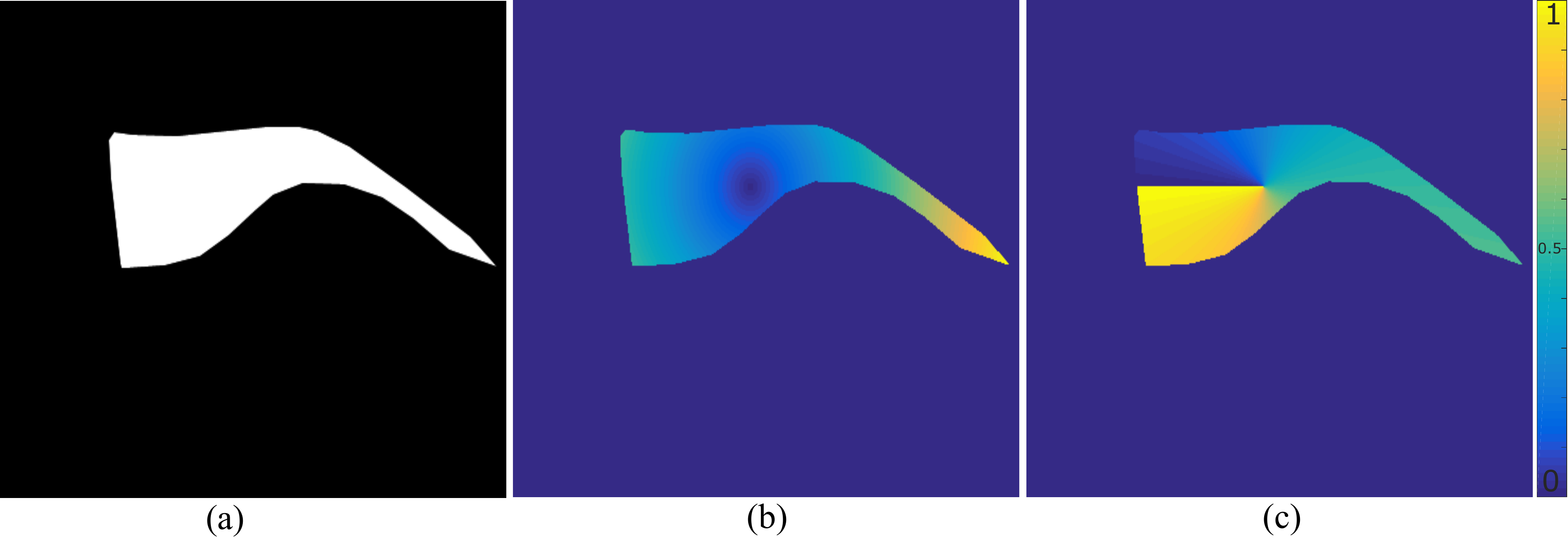}
	\caption{The proposed positional information (PI) maps to be used as auxiliary inputs for the tendinopathy recognition task. (a) shows the original mask of the tendon region, while (b) and (c) images illustrate the proposed normalized radius (${{\bf{r}}_{\mathcal{N}}}$)  and angle (${{\bf{\theta}}_{\mathcal{N}}}$) maps obtained based on it, respectively.}
\label{fPosMaps}
\end{figure*}

Positioning maps $(\bf{r},\Theta)$ are standardized to be robust against the changes in the object scale and rotation by applying the following transformation:
\begin{equation}
    {{\cal N} ({\cal X}) = {{({\cal X} - \min ({\cal X}))} \mathord{\left/
 {\vphantom {{({\cal X} - \min ({\cal X}))} {\max({\cal X})}}} \right.
 \kern-\nulldelimiterspace} {(\max({\cal X})-\min({\cal X})})}},
\end{equation}
in the above equation, $\mathcal{X}$ can be replaced with any of ${\Theta}$ or $\bf{r}$ to achieve normalized positioning maps, ${{{\Theta}}_{\mathcal{N}}}$ and ${{\bf{r}}_{\mathcal{N}}}$. This function transfers the values of each map to $(0,1)$ closed interval. Note that this effect will also make the positioning maps’ values more favorable for network training. A sample tendon mask is illustrated in \cref{fPosMaps} alongside its constructed normalized radius (${{\bf{r}}_{\mathcal{N}}}$) and angle (${{\bf{\theta}}_{\mathcal{N}}}$) maps.

In the training procedure of tendinopathy recognition network, these two positioning maps are concatenated with the original US image to be fed into the network as a three channeled input, like the illustration in \cref{fRecogArch}. By doing this, the network can capture the spatial information alongside the visual features of the US image. This spatial awareness property is conceivable via residual and Skip connections in the normal cell, which make it able to carry the positional information (PI) throughout the network \cite{alemi2019spanet}.

\subsubsection{Network architecture}
\label{sec:clsNet}
\begin{figure*}[htb]
	\centering
    \includegraphics[width=1\textwidth]{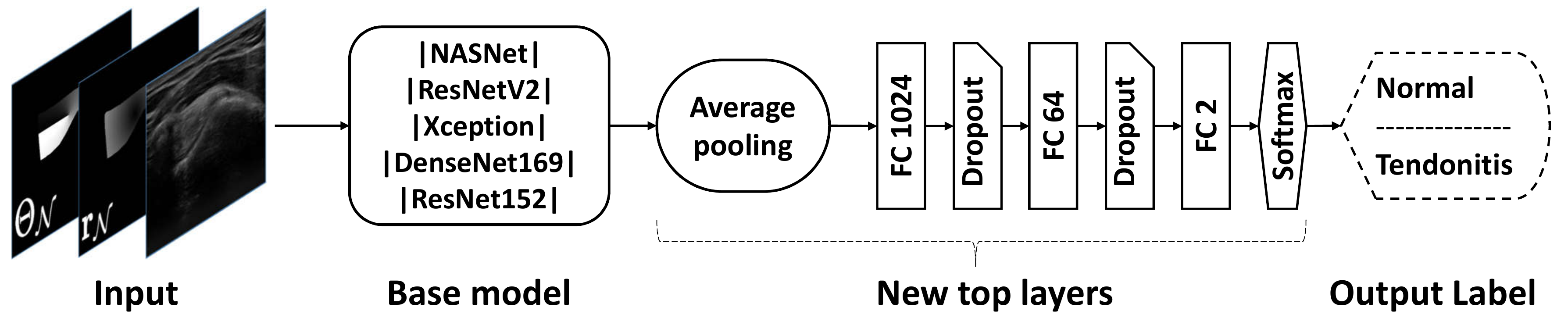}
	\caption{The proposed tendinopathy recognition network. This network accepts the US image alongside with normalized radius (${{\bf{r}}_{\mathcal{N}}}$)  and angle (${{\bf{\theta}}_{\mathcal{N}}}$) maps as the input. Top layers of  base models must be eliminated in order to add new top layers.}
\label{fRecogArch}
\end{figure*}

A general classification pipeline has been designed for tendinopathy recognition, as illustrated in \cref{fRecogArch}. The whole network consists of a base model (backbone), and new top layers sequentially added to it. The backbone act as the recognition engine and carries the duty of relevant feature generation. Those features are then used and processed through top layers to construct the final prediction. In the classification framework of \cref{fRecogArch}, the base model can be any of NASNet \cite{nasnet}, Inception-ResNet v2 (ResNetV2) \cite{resnetv2}, Xception \cite{xception}, 169-layered DenseNet (DenseNet169) \cite{densenet}, or 152-layered ResNet \cite{resnet} as long as their original top layers are eliminated. However, the recognition framework in this research is designed to work optimally with NASNet-A model (illustrated in the encoding part NASUNet in \cref{fSegArch}).

Our proposed top layers consist of one average pooling layer to reduce the size of feature maps and completely flatten them, two dense (fully connected) layers with 1024 and 64 nodes each followed by Dropout layer, and finally a two-node fully connected layer that serves as the decision layer of the recognition network. The final prediction is generated by applying Softmax \cite{lecun2015deep} on the output of the last layer.

\subsubsection{Training procedure}
The proposed classification network in \cref{fRecogArch} includes two main parts: the base model and top layers. The base model kernels are initialized with pre-trained ImageNet weights, whereas top layers are initialized randomly. Just like the segmentation network, for the classification network, our proposed two-phase transfer learning paradigm is carried out as well. First, the recognition network is trained on the UNS data set, and then it is fine-tuned on the available tendinopathy data set. For the first phase, the UNS data set is transferred into a two-category classification data set by labeling the images with the presence of Brachial Plexus nerve structures as the positive class and assigning the other images to the negative class.

As one can see in \cref{fRecogArch}, the network takes a three channeled input containing US image and positional information maps. Positional information maps were previously normalized, ergo, we normalize the US image by transforming its 8-bit integer values into 32-bit floating point scalars in the range of 0 to 1. This normalization will make the US image values suitable for network training and more coincide with the other two channels in the input. The network weights are optimized by Adam optimizer \cite{kingma2014adam}. In the first phase of training, learning rates of 0.001 and 0.005 are utilized for training the base model and top layers of the classification network, respectively. However, in the second phase, the learning rate of 0.0005 is selected for all network layers. All recognition experiments are done in a 5-fold cross-validation framework, and each training procedure lasts for 200 epochs. 

\section{Experimental setup and results}
\label{sec:results}

\subsection{Evaluation metrics}
Two binary image similarity criteria have been used to evaluate segmentation performance, Dice similarity coefficient (DSC) and Jaccard similarity index (JSI). If $\bf{G}$ be the binary mask of the ground truth and $\bf{P}$ be the prediction segmentation mask resulted from the proposed segmentation method, two similarity criteria as mentioned earlier can be defined as below:
\begin{itemize}
    \item $ JSI = {{\left| {{\bf{G}} \cap {\bf{P}}} \right|} \mathord{\left/
 {\vphantom {{\left| {{\bf{G}} \cap {\bf{P}}} \right|} {\left| {{\bf{G}} \cup {\bf{P}}} \right|}}} \right.
 \kern-\nulldelimiterspace} {\left| {{\bf{G}} \cup {\bf{P}}} \right|}} $
 
    \item $ DSC = 2{{\left| {{\bf{G}} \cap {\bf{P}}} \right|} \mathord{\left/
 {\vphantom {{\left| {{\bf{G}} \cap {\bf{P}}} \right|} {\left( {\left| {\bf{G}} \right| + \left| {\bf{P}} \right|} \right)}}} \right.
 \kern-\nulldelimiterspace} {\left( {\left| {\bf{G}} \right| + \left| {\bf{P}} \right|} \right)}} $
\end{itemize}

Besides these two similarity indices, three pixel-based segmentation evaluation metrics are reported in this research. These metrics are defined based on considering the image segmentation as a dense pixel classification task.

The second group of metrics is defined based on the number of true positives (TP:= \#correctly predicted as positive sample/pixel), false positives (FP:= \#falsely predicted as positive sample/pixel), true negative (TN:= \#correctly predicted as negative sample/pixel), and false negatives (FN:= \#falsely predicted as negative sample/pixel). These evaluation metrics are accuracy (ACC), sensitivity (SEN), and specificity (SPC):
\begin{itemize}
   \item $ ACC = {{\left( {{\rm{TP + TN}}} \right)} \mathord{\left/
 {\vphantom {{\left( {{\rm{TP + TN}}} \right)} {\left( {{\rm{TP + FP + TN + FN}}} \right)}}} \right.
 \kern-\nulldelimiterspace} {\left( {{\rm{TP + FP + TN + FN}}} \right)}} $
 
    \item $ SEN = {{{\rm{TP}}} \mathord{\left/
 {\vphantom {{{\rm{TP}}} {\left( {{\rm{TP + FN}}} \right)}}} \right.
 \kern-\nulldelimiterspace} {\left( {{\rm{TP + FN}}} \right)}} $

\item $ SPC = {{{\rm{TN}}} \mathord{\left/
 {\vphantom {{{\rm{TN}}} {\left( {{\rm{TN + FP}}} \right)}}} \right.
 \kern-\nulldelimiterspace} {\left( {{\rm{TN + FP}}} \right)}} $
\end{itemize}
These three criteria are common between classification and segmentation tasks. Furthermore, there are another two metrics used for classification evaluation: 1) Precision or positive predictive value (PPV), defined as below:
\begin{itemize}
       \item $ PPV = {{{\rm{TP}}} \mathord{\left/
 {\vphantom {{{\rm{TP}}} {\left( {{\rm{TP + FP}}} \right)}}} \right.
 \kern-\nulldelimiterspace} {\left( {{\rm{TP + FP}}} \right)}} $
\end{itemize}
2) Area Under the Curve (AUC) which is the area between the receiver operating characteristics (ROC) curve and its horizontal axis \cite{ROC}. For classification task (and the segmentation task if be considered as pixel classification problem \cite{jahanifar2019supervised,jahanifar2018segmentation}), ROC curve is generated by plotting the true positive prediction rate against the false positive prediction rate at different threshold settings applied on the predictions (which can be class probability or pixel values prediction) \cite{ROC}.

Each criterion defined above has its concept and interpretation, which explaining them is beyond the scope of the current paper. Nevertheless, all criteria (DSC, JSI, ACC, SEN, SPC, AUC, PPV) have their possible values between 0 and 1. The higher the metrics, the better segmentation, or classification performance will be. For better representation,  these metrics are scaled into percentages in the reports. In ROC curves, each curve that stands above the others indicates better performance and results in higher AUC values \cite{ROC}.

\subsection{Implementation details}
All training and evaluation experiments have been done in Python using Keras (with Tensorflow backend) framework on an Intel Core i7 workstation equipped with two GPUs (NVidia 1080 Ti) and 64 GB of RAM.

\subsection{Segmentation results}

Resulted segmentations of different models are evaluated using accuracy, Dice, Jaccard, sensitivity, and specificity metrics, which are reported in \cref{tSegRes}. Reported metrics are the average values over all images of SST data-set. As one can see, our proposed segmentation model, NASUNet,  outperform other segmentation networks with a high margin. Obtaining the values of 95.21\%, 93.85\%, 87.32\%, 93.26\%, and 96.33\% for ACC, DSC, JSI, SEN, and SPC metrics, respectively,  NASUNet model shows a boost in performance on SST data set. Compared to the second best performing model, DeepLab V3, our proposed method outperforms by a high margin of 2.11\% and 1.46\% for DSC and JSI metrics, respectively. 

\begin{table}[]
\centering
\begin{tabular}{lccccc}
\hline \hline
Model              & ACC   & DSC   & JSI   & SEN   & SPC   \\
\hline
NASUNet (proposed) & 95.21 & 93.85 & 87.32 & 93.26 & 96.33 \\
DeepLab v3 \cite{deeplab}         & 94.15 & 91.74 & 85.86 & 92.17 & 95.30 \\
SegNet \cite{segnet}           & 93.88 & 91.02 & 83.90 & 90.17 & 96.45 \\
U-Net \cite{unet}           & 94.55 & 88.70 & 82.09 & 90.28 & 97.21 \\
FCN-8 \cite{fcn}          & 93.08 & 85.12 & 78.38 & 87.41 & 95.24 \\
AAM \cite{cootes2001active}            & 88.40 & 81.31 & 74.88 & 85.38 & 91.22 \\
\hline\hline
\end{tabular}
\caption{Average of resulted segmentation evaluation metrics over the SST data set reported for different segmentation models}
\label{tSegRes}
\end{table}

\begin{figure*}[htb]
	\centering
    \includegraphics[width=0.6\textwidth]{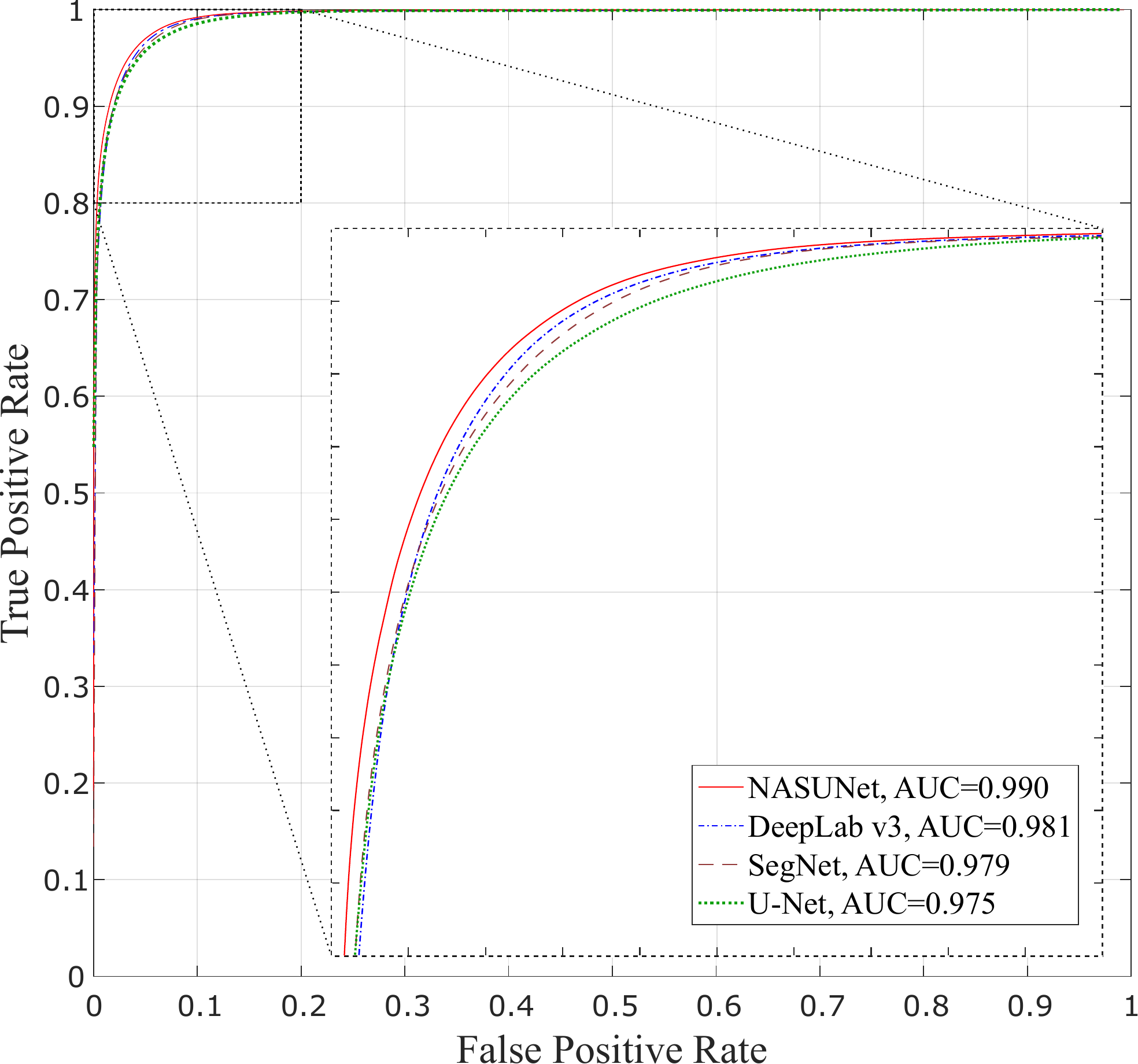}
	\caption{ROC curves and AUC values resulted from evaluating different segmentation models on the SST data set. The inner figure represents a zoomed view of the annotated area, where ${\rm{FPR}} \in (0,0.2)$ and ${\rm{TPR}} \in (0.8,1)$.}
\label{fSegROC}
\end{figure*}

The ROC curve for NASUNet in  \cref{fSegROC} stands above AUC curves of other methods resulting in higher AUC value (\cref{fSegROC}). NASUNet acheives AUC value of 0.99 for  against 0.981, 0.979, and 0.975 AUC values achieved for DeepLab v3 \cite{deeplab}, SegNet \cite{segnet}, and U-Net \cite{unet} models, respectively.

\begin{figure}[htb]
    \centering 
\begin{subfigure}{0.25\textwidth}
  \includegraphics[trim={0 150 0 0},clip, width=\linewidth]{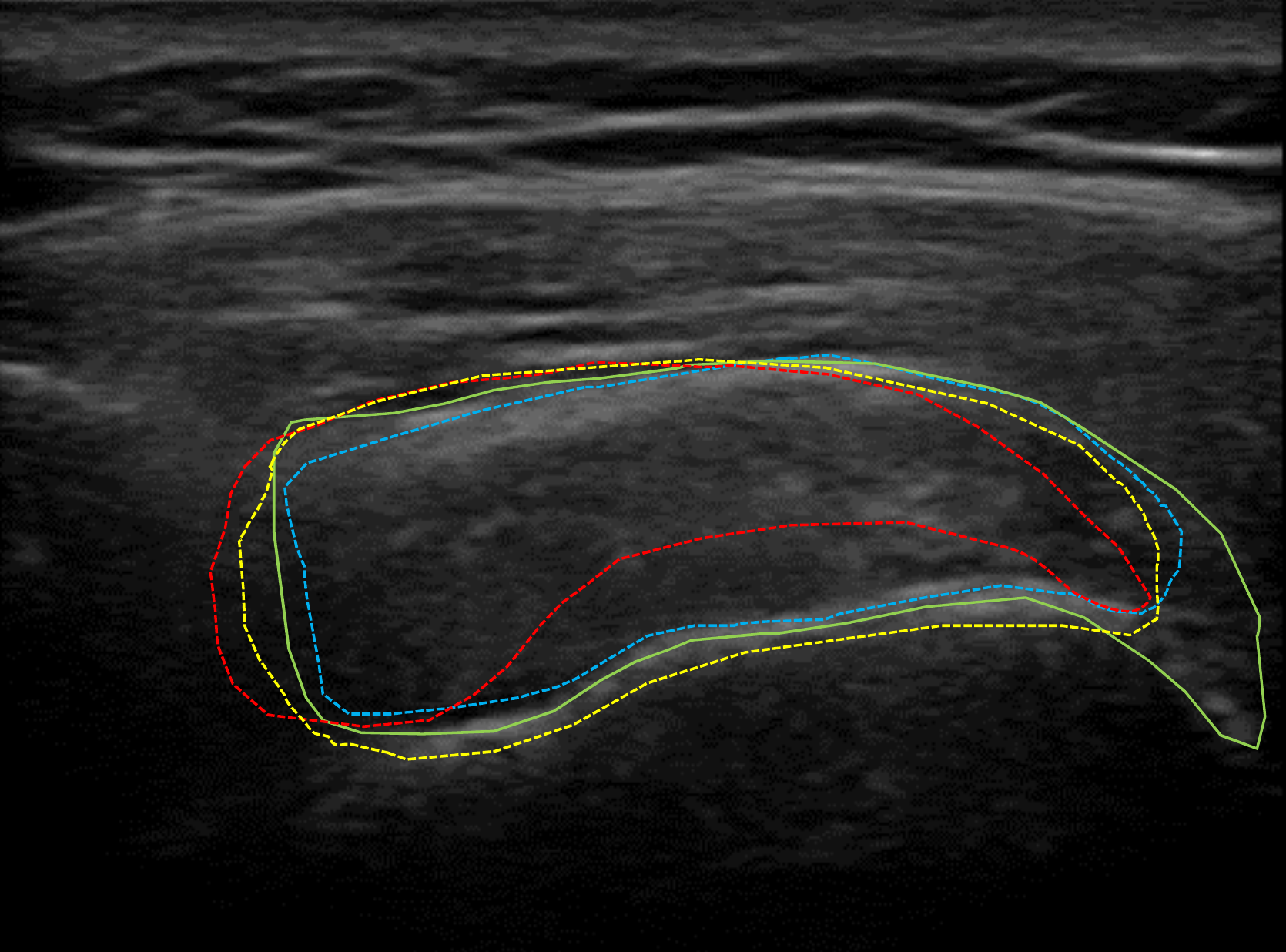}
\end{subfigure}\hfil 
\begin{subfigure}{0.25\textwidth}
  \includegraphics[trim={0 150 0 0},clip, width=\linewidth]{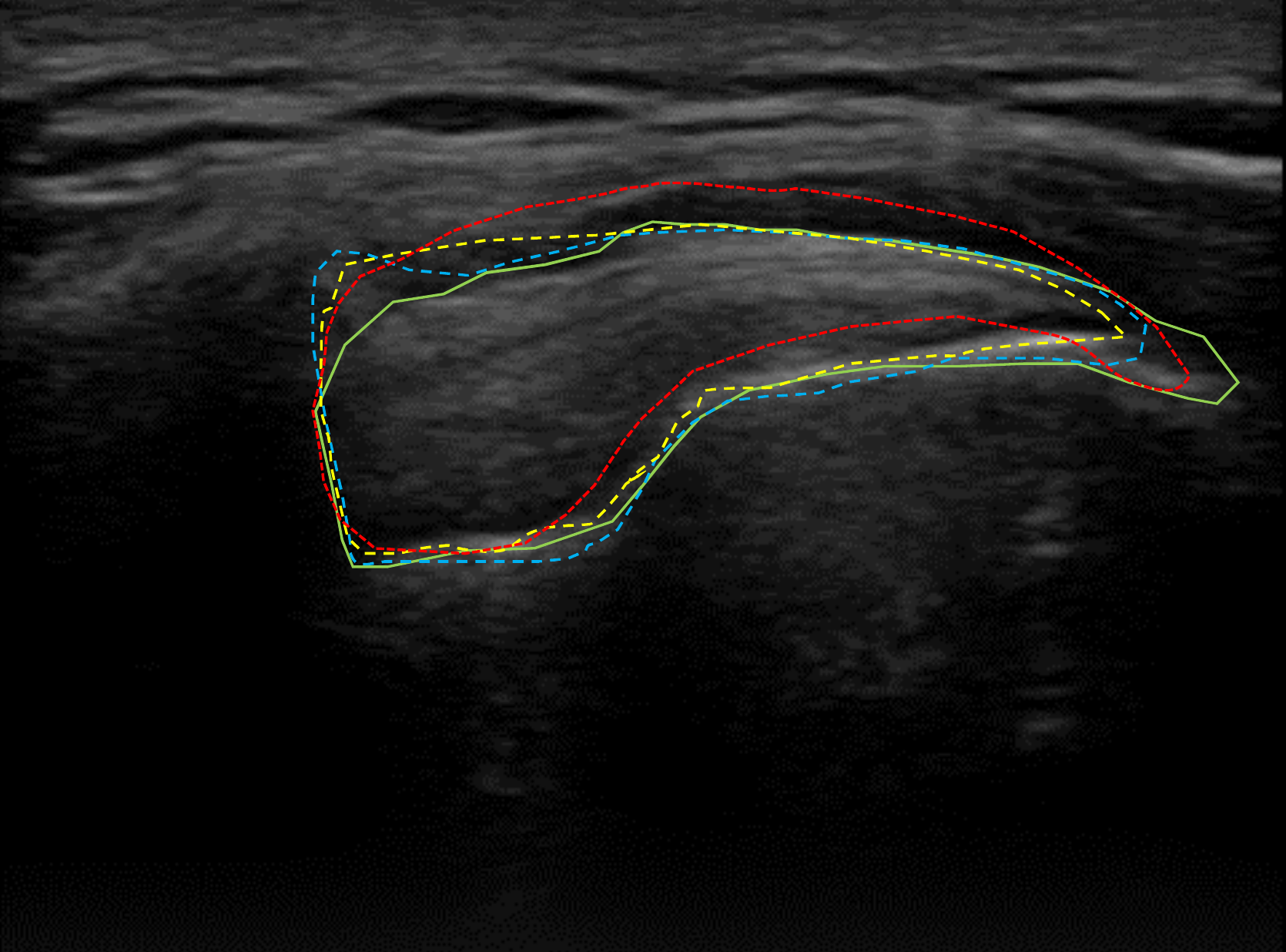}
\end{subfigure}\hfil 
\begin{subfigure}{0.25\textwidth}
  \includegraphics[trim={0 150 0 0},clip, width=\linewidth]{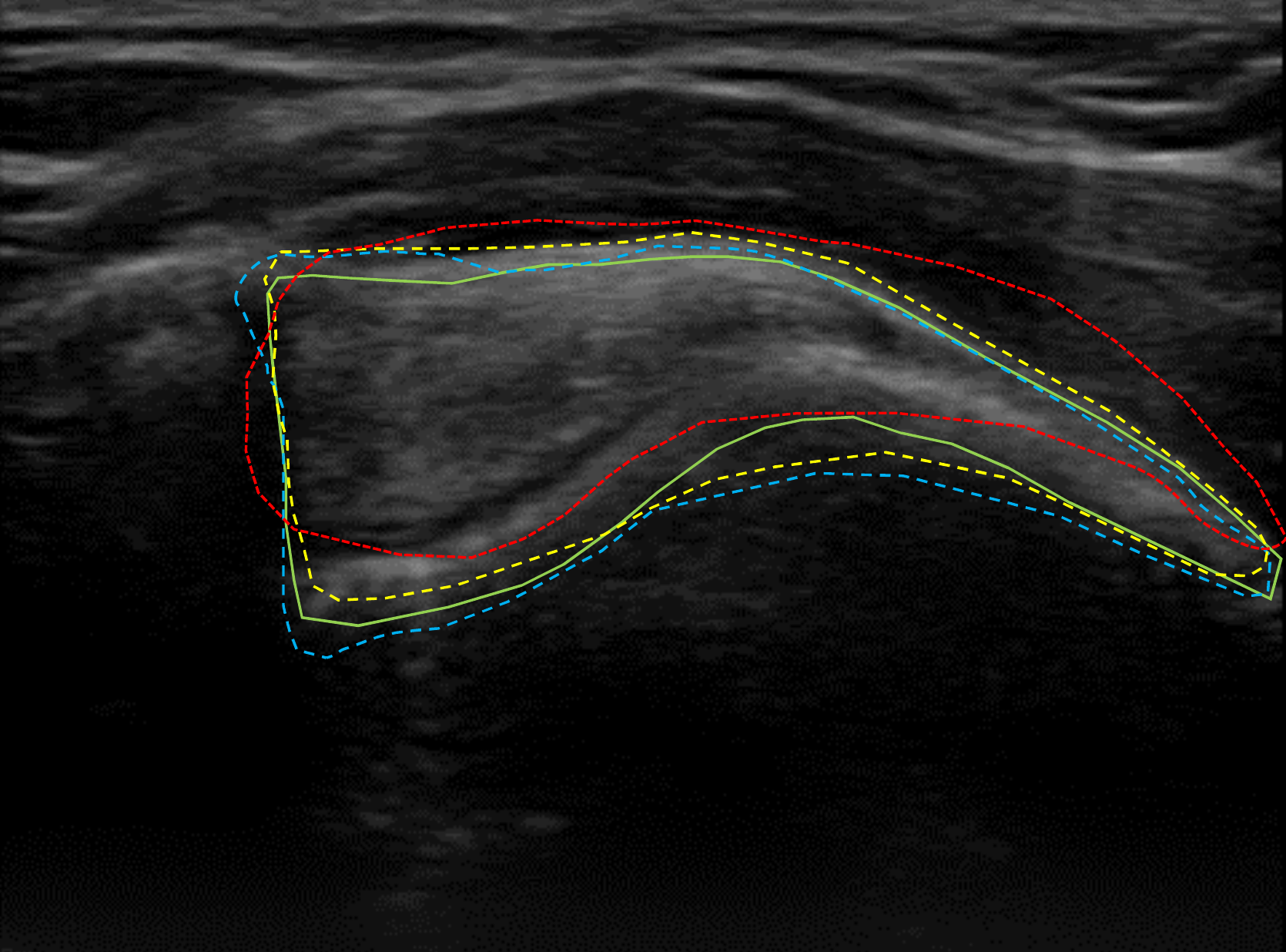}
\end{subfigure}

\medskip
\begin{subfigure}{0.25\textwidth}
  \includegraphics[trim={0 150 0 0},clip, width=\linewidth]{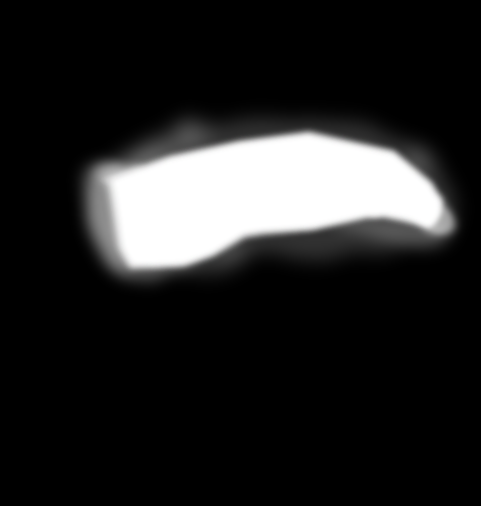}
\end{subfigure}\hfil 
\begin{subfigure}{0.25\textwidth}
  \includegraphics[trim={0 150 0 0},clip,width=\linewidth]{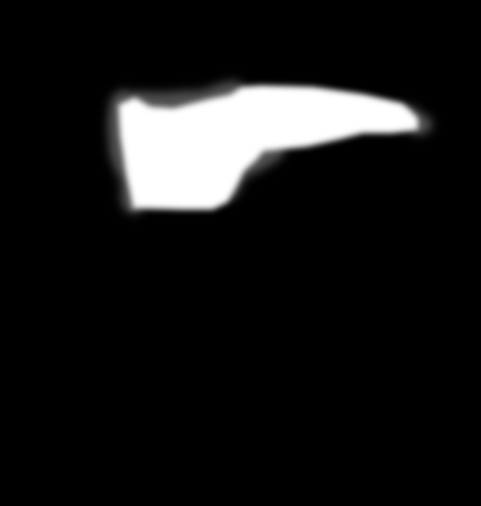}
\end{subfigure}\hfil 
\begin{subfigure}{0.25\textwidth}
  \includegraphics[trim={0 150 0 0},clip,width=\linewidth]{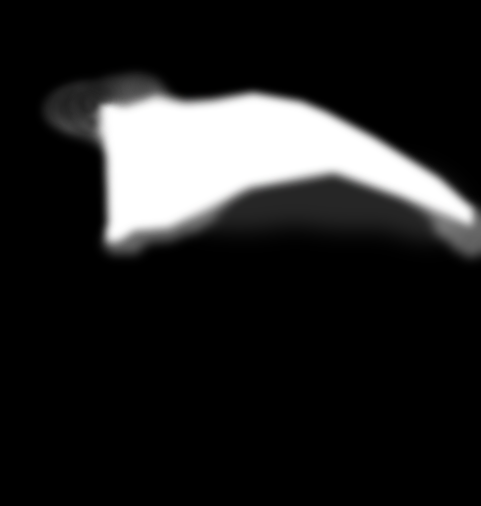}
\end{subfigure}
\caption{Visualizing the SST boundaries resulted from different segmentation methods for three different cases in the first row. Boundaries of ground truth, proposed NASUNet model, DeepLab v3 network, and AAM method are overlaid on their related images with green, blue, yellow, and red colors, respectively. The second row illustrates the raw segmentation prediction maps from NASUNet.}
\label{fSegRes}
\end{figure}

Qualitative performance and advantage of our proposed segmentation method over two different methods are presented in \cref{fSegRes}. In this figure, original US images of three cases taken from the SST data-set are shown in the first row and the second row depicts the raw predictions of the NASUNet model. On each image, boundaries of the  segmentation maps derived from the ground truth (radiologist annotation), proposed method (NASUNet), DeepLab V3 network (the second best performing method in \cref{tSegRes}), and Active Appearance Model (AAM, the worst-performing method based on \cref{tSegRes}) are drawn with green, blue, yellow, and red colors, respectively. It is evident from \cref{fSegRes} that the delineation by our method is closer to the ground truth annotation in comparison with other methods, which indicates the power and efficiency of NASUNet in SST boundary detection. In all of the subsequent tendinopathy recognition experiments, resulting segmentations from the best performing model (NASUNet) are utilized.

\subsection{Recognition results}
Three different scenarios for SST classification have been tested. All scenarios share the same classification network architecture pipeline (explained in \cref{sec:clsNet}), loss function, and training procedures, but, the input is different for each scenario. In other words, three different strategies are Incorporated to provide the input for the classification network. The first scenario uses only the gray-scale US image in the input (OI strategy), the second scenario incorporates the tendon mask alongside the US image in the input (WM strategy), and in the third scenario US image is concatenated with two positional information map (PI strategy as described in \cref{sec:posInf}). Each scenario is tested on the classification of SST data set in 5-fold cross-validation experiments. In our experiments, the base model in the classification network (\cref{fRecogArch}) is replaced with five powerful CNN models interchangeably.

For the proposed classification framework with three different input strategies, results related to  OI, WM, and PI scenarios are reported in \crefrange{tClsRes_WOM}{tClsRes_PI}, respectively. As shown, in all scenarios, the NASNet-PI model shows better performance compared to other models in the classification framework (see \cref{tClsRes_NASNet}). For the best performing scenario (using positional information) as reported in \cref{tClsRes_PI}, NASNet-PI model can achieve values of 91.00\%, 86.67\%, and 92.86\% for ACC, SEN, and SPC metrics, respectively. Comparing these results to the values for ResNet152-OI model in the worst performing scenario as reported in \cref{tClsRes_WOM} (ACC of 71\%, SEN 60\%, and SPC of 75.71\% in tendon classification), a great improvement in the recognition performance can be seen (ACC, SEN, and SPC metrics have increased in amount of 31\%, 26.67\%, and 17.15\%, respectively), which is due to incorporating the knowledge transferring concept (by using NASNet architecture) and the proposed positional information maps in the input. However, even in the scenario with positional information, the second-best  model (DenseNet169-PI which gains ACC of 88.00\%, SEN of 83.33\%, and SPC of 90.00\%) underperforms the NASNet-PI according to all criteria, i.e., the NASNet-PI model surpasses the DenseNet169-PI by 3\%, 3.34\%, and 2.86\% increment in ACC, SEN, and SPC metrics, respectively. 

\begin{table}[h]
\centering
\begin{tabular}{lccccc}
\hline \hline
\textbf{Base model} & \textbf{AUC} & \textbf{ACC} & \textbf{SEN} & \textbf{SPC} & \textbf{PPV} \\
\hline
NASNet         & 84.62        & 77.00        & 73.33        & 78.57        & 59.46        \\
ResNetV2       & 73.05        & 74.00        & 66.67        & 77.14        & 55.56        \\
Xception       & 82.05        & 73.00        & 66.67        & 75.71        & 54.05        \\
DenseNet169    & 67.48        & 73.00        & 63.33        & 77.14        & 54.29        \\
ResNet152      & 68.67        & 71.00        & 60.00        & 75.71        & 51.43   \\
\hline\hline
\end{tabular}
\caption{OI scenario: Classification evaluation metrics reported for SST data set resulted from different base models in the proposed recognition pipeline, without including any of SST mask or positional information in the input.}
\label{tClsRes_WOM}
\end{table}

\begin{table}[h]
\centering
\begin{tabular}{lccccc}
\hline \hline
\textbf{Base model} & \textbf{AUC} & \textbf{ACC} & \textbf{SEN} & \textbf{SPC} & \textbf{PPV} \\
\hline
NASNet         & 93.81        & 88.00        & 80.00        & 91.43        & 80.00        \\
ResNetV2       & 92.67        & 86.00        & 76.67        & 90.00        & 76.67        \\
Xception       & 92.10        & 85.00        & 73.33        & 90.00        & 75.86        \\
DenseNet169    & 86.14        & 85.00        & 76.67        & 88.57        & 74.19        \\
ResNet152      & 85.14        & 83.00        & 70.00        & 88.57        & 72.41   \\
\hline\hline
\end{tabular}
\caption{WM scenario: Classification evaluation metrics reported for SST data set resulted from different base models in the proposed recognition pipeline, including SST mask in the input.}
\label{tClsRes_WM}
\end{table}

\begin{table}[h]
\centering
\begin{tabular}{lccccc}
\hline \hline
\textbf{Base model} & \textbf{AUC} & \textbf{ACC} & \textbf{SEN} & \textbf{SPC} & \textbf{PPV} \\
\hline
NASNet         & 96.24        & 91.00        & 86.67        & 92.86        & 83.87        \\
ResNetV2       & 95.71        & 88.00        & 83.33        & 90.00        & 78.13        \\
Xception       & 92.29        & 87.00        & 76.67        & 91.43        & 79.31        \\
DenseNet169    & 95.24        & 89.00        & 83.33        & 91.43        & 80.65        \\
ResNet152      & 89.29        & 85.00        & 73.33        & 90.00        & 75.86      \\
\hline\hline
\end{tabular}
\caption{PI scenario: Classification evaluation metrics reported for SST data set resulted from different base models in the proposed recognition pipeline, including SST positional information in the input.}
\label{tClsRes_PI}
\end{table}

To better illustrate the effect of using each scenario, in \cref{fClsROC} and \cref{tClsRes_NASNet} the ROC curves and classification performance metrics related to the NASNet-based classification model’s outputs in different scenarios are reported.

\section{Discussion}
\label{sec:disscution}
Two main challenges of automatic recognition of tendinopathy from US images are data shortage and indistinct appearance of US images. The problem of the small data set poses serious hurdles when approaching deep neural networks solutions. To work around this problem, concepts of knowledge transferring, transfer learning, and data augmentation are incorporated. Relying on knowledge transferring,  our network architecture is built based on pre-designed constructing blocks \cite{nasnet}, and with transfer learning network weights that were pre-trained on another big-enough related data set \cite{van2015transfer} are used. These two concepts helped us deal with the common problem of data shortage in medical applications, properly.

\subsection{Tendon segmentation task}
Based on \cref{tSegRes}, our proposed segmentation network (NASUNet) can outperform all other well-known segmentation networks. However, in order to discuss the strength of this method and the effect of different items on it, two other segmentation experiments are arranged. In one scenario, the NASUNet network is tested on SST data set segmentation task without pre-training it on the UNS data-set while using DATs (NASUNet Augmented), and in another scenario, a vanilla NASUNet was trained on SST data set with no pre-training phase nor incorporating DATs (NASUNet Vanilla). Results of this experiment are reported in \cref{tSegRes_nasunet}. This experiment will help us measure the effect of transfer learning and data augmentation techniques on the segmentation task.

\begin{table}[]
\centering
\begin{tabular}{lccc}
\hline \hline
Model              & Description   & DSC   & JSI   \\
\hline
NASUNet Pre-trained & Both pre-training and DAT incorporated & 93.85 & 87.32 \\
NASUNet Augmented & Only DAT incorporated & 92.81 & 86.91\\
NASUNet Vainlla  & No pre-training nor DAT is incorporated & 89.11 & 82.35 \\
\hline\hline
\end{tabular}
\caption{Performance evaluation of NASUNet segmentation model on SST data set for different scenarios of pre-training and data augmentation techniques (DAT).}
\label{tSegRes_nasunet}
\end{table}

By comparing the results of \cref{tSegRes} with "NASUNet Augmented" performance in \cref{tSegRes_nasunet}, it is evident that even without considering the transfer learning our proposed segmentation network can outperform the second-best model (DeepLabV3) by 1.07\% and 1.05\%  for DSC and JSI metrics, respectively. This precedence shows the efficiency of the proposed segmentation architecture empowered by knowledge transferring over other state-of-the-art networks. Segmentation performance gets even better when transfer learning is incorporated (NASUNet Pre-trained), i.e., when our proposed segmentation network is first pre-trained on a more comprehensive data set of US images (UNS data set \cite{uns}) and then fine-tuned on the available tendon data set. In this mode, our proposed segmentation network outperforms other approaches with no transfer learning by a large margin, showing improvement of 1.04\% and 0.41\%  for DSC and JSI. That is due to a large number of training data, allowing the network to learn more relevant feature kernels and have high generalizability after fine-tuning.

Data augmentation is another technique to deal with data shortage. 
The positive effect of DATs has been proven in the literature \cite{jahanifar2018segmentation}. However, for comparison purposes, in \cref{tSegRes_nasunet}, the segmentation results obtained from NASUNet without DATs and pre-training (NASUNet Vanilla) are reported. As expected, "NASUNEt Vanilla" performed considerably worse than the "NASUNet Augmented" and "NASUNet Pre-trained." For example, only incorporating DATs makes a significant advance of 3.7\% and 4.56\% in DSC and JSI, respectively, which approves the high impact of DATs on model generalization. 

\subsection{Tendinopathy recognition task}
The effect of the proposed workaround for the problem of intricate texture appearance and noisy nature of US images is discussed here. In  \cref{sec:posInf} we proposed to provide the classification network with positioning information in the form of two auxiliary input channels to the network. As depicted in \cref{tClsRes_PI}, our proposed method outperforms all other state-of-the-art models. As \crefrange{tClsRes_WOM}{tClsRes_PI} shows, NASNet base network in the proposed classification pipeline achieves better results compared to other well-known classification networks, for all network input scenarios, which is because of knowledge transferring and extensive use of multi-scale convolutional blocks, empowering the architecture of NASNet to extract more relevant feature for tendon classification. 

\begin{figure*}[!h]
	\centering
    \includegraphics[width=0.6\textwidth]{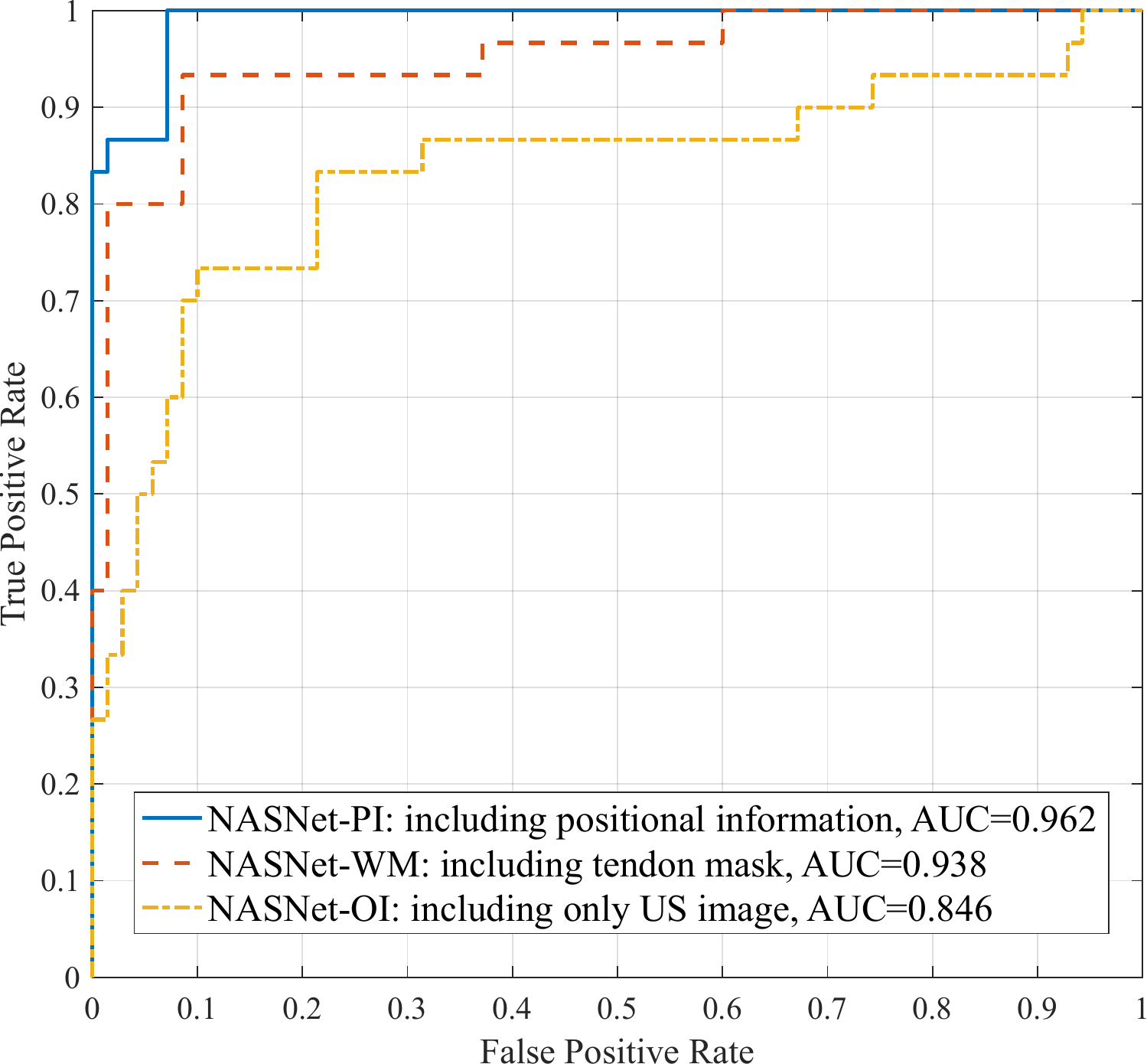}
	\caption{ROC curves and AUC values resulted from evaluating NASNet base model in the proposed tendon classification framework on the SST data set, using three different input strategies (scenarios): NASNet-OI, NASNet-WM, and NASNet-PI.}
\label{fClsROC}
\end{figure*}

However, as reported in \cref{tClsRes_NASNet} and depicted in ROC curves of \cref{fClsROC}, the performance of NASNet is considerably elevated when tendon boundary information is added to the network input (NASNet-WM in comparison with NASNet-OI).
When using NASNet base model in the classification framework, adding the mask information (NASNet-WM) increased the accuracy, sensitivity, and precision of tendinopathy recognition  11\%, 6.67\%, and 20.54\% compared to NASNet-OI, respectively. These metrics are further increased when positional information are taking into account (NASNet-PI), reaching values of 91\%, 86.67\%, and 83.87\% for recognition accuracy, sensitivity, and precision. 24.41\% improvement in classification precision for the NASNet-PI over the NASNet-OI model shows the distinction power of the proposed classification method incorporating positional information. According to \crefrange{tClsRes_WOM}{tClsRes_PI} same trend can be found for other base models, where incorporating positional information elevates the results outstandingly. The superiority of the proposed NASNet-PI over NASNet-OI and NASNet-WM is also apparent from ROC curves of \cref{fClsROC}, in which NASNet-PI stands above all, achieving AUC of 0.962 against AUC of 0.938 for NASNet-WM and AUC of 0.846 for NASNet-OI.

\begin{table}[]
\centering
\begin{tabular}{lccccc}
\hline \hline
\textbf{Base model} & \textbf{AUC} & \textbf{ACC} & \textbf{SEN} & \textbf{SPC} & \textbf{PPV} \\
\hline
NASNet-PI        & 96.24        & 91.00        & 86.67        & 92.86        & 83.87        \\
NASNet-WM        & 93.81        & 88.00        & 80.00        & 91.43        & 80.00        \\
NASNet-OI         & 84.62        & 77.00        & 73.33        & 78.57        & 59.46        \\
\hline\hline
\end{tabular}
\caption{Classification evaluation metrics reported for SST data set resulted from NASNet base model in the proposed recognition pipeline, using three different input scenarios.}
\label{tClsRes_NASNet}
\end{table}

Nature of the NSANet constructing blocks (Normal and Reducing cells) enables feature map aggregation through the network path. As seen in \cref{fNormReducBlock}, alongside the modified feature maps (output of convolutional and average pooling layers), Normal cell has a Skip connection that transfers the  $h_{i-1}$ to the output as-is and concatenates it with the outputs of other layers. Feature aggregation enables the network harvesting the positional information (auxiliary inputs) in every stage of its path, making it a spatial aware network. Dense blocks in DensNet169 architecture \cite{densenet} also show similar feature aggregation behavior which leads to use lesion mask and positional information in \cref{tClsRes_WM,tClsRes_PI} and gain better performance in comparison with ResNetV2, Xception, and ResNet169.

Intricate textural patterns that indicate tendinopathy or tendon inflammation are very similar to the appearance of the image in the irrelevant areas.
To bring more clarity to this concept, take \cref{fSST2} into consideration, in which a tendinopathy sample is illustrated, and the injured regions are annotated with orange color. The appearance of the injured area is very similar to textural structures outside the tendon region. 
Therefore,  network awareness of tendon region (like NASNet-WM) can perform considerably better than ordinary models (like NASNet-OI). Furthermore, from radiology aspect, it is crucial to know the location where a specific textural pattern occurs. For example, injury markers in supraspinatus tendon are usually occurring in its peripheral regions \cite{lipman2018}. Therefore, providing the network with the relative position of textural structures inside the tendon region can increase the accuracy of diagnosis, as for the NASNet-PI in \cref{tClsRes_NASNet} and \cref{fClsROC}.

When comparing our recognition results with reported results in other extensive systematic reviews that studied US and MRI interpretations against surgical findings \cite{roy2015diagnostic, saraya2016}, performance measures are very similar. This comparison shows that our results are propitious, and the proposed method can help designing better treatment planning, patient management, and realizing precision medicine.
The main drawback of the current research is the lack of training/validation data. We have tried to overcome this problem by leveraging transfer learning and data augmentation techniques.  High evaluation metrics reported from cross-validation experiments approve our policy facing the problem. However, the best practice in this area is to extend the data set. A significant amount of data can guarantee training a more generalizable model and avoiding over-fitting \cite{lecun2015deep,litjens2017survey}. Another useful practice is to enrich the data set by manually segmenting the injured area in the tendinopathy cases. This way, the tendinopathy recognition problem can be exploited as a dense prediction (segmentation) or a detection problem.

\section{Conclusion}
\label{sec:conclusion}
In this paper, the focus is on the automatic recognition of tendinopathy from US images. To this end, an exclusive data set of labeled US image from the supraspinatus tendon is acquired. Main challenges are the complex textural pattern and noisy appearance of US images and lack of data. To overcome these hurdles, the problem is divided into two separate tasks of tendon boundary segmentation and disease classification. The knowledge transferring, transfer learning, and data augmentation techniques are employed in network training. A novel segmentation network architecture is proposed based on constructing blocks of NASNet model (Normal and Reducing cells) and another novel constructing block, named Enlarging cell. A two-component loss function is used for segmentation task based on modified Dice similarity coefficient and cross-entropy. The segmentation network can outperform other well-known methods in the literature, achieving DSC and JSI values of 93.85\% and 87.32\% in the cross-validation experiments, respectively. For the classification task, the modified NASNet-A (6 @ 4032) model \cite{nasnet} is fed with positional information of tendon region alongside original US image (NASNet-PI). At this moment, we can gain better recognition accuracy of 91\%, sensitivity of 86.67\%, and specificity of 92.86\%  compared with other state-of-the-art classification models.

\section*{References}

\bibliography{mybibfile}

\begin{thebibliography}{10}
\expandafter\ifx\csname url\endcsname\relax
  \def\url#1{\texttt{#1}}\fi
\expandafter\ifx\csname urlprefix\endcsname\relax\def\urlprefix{URL }\fi
\expandafter\ifx\csname href\endcsname\relax
  \def\href#1#2{#2} \def\path#1{#1}\fi

\bibitem{lipman2018}
K.~Lipman, C.~Wang, K.~Ting, C.~Soo, Z.~Zheng, Tendinopathy: injury, repair,
  and current exploration, Drug Design, Development and Therapy 12 (2018)
  591--603.

\bibitem{sharma2006biology}
P.~Sharma, N.~Maffulli, Biology of tendon injury: healing, modeling and
  remodeling, Journal of Musculoskeletal and Neuronal Interactions 6~(2) (2006)
  181--190.

\bibitem{de2018ultrasound}
J.~de~la Fuente, M.~Blasi, S.~Mart{\'\i}nez, P.~Barcel{\'o}, C.~Cach{\'a}n,
  M.~Miguel, C.~Pedret, Ultrasound classification of traumatic distal biceps
  brachii tendon injuries, Skeletal Radiology 47~(4) (2018) 519--532.

\bibitem{chang2009imaging}
A.~Chang, T.~T. Miller, Imaging of tendons, Sports Health 1~(4) (2009)
  293--300.

\bibitem{stevic2013us}
R.~Stevic, M.~Dodic, Ultrasonography of tendon abnormalities, OA
  Musculoskeletal Medicine 1~(2) (2013) 12--19.

\bibitem{saraya2016}
S.~Saraya, R.~E. Bakry, Ultrasound: Can it replace mri in the evaluation of the
  rotator cuff tears?, The Egyptian Journal of Radiology and Nuclear Medicine
  47~(1) (2016) 193--201.

\bibitem{scott2018diagnostic}
A.~Scott, T.~M. Zaharadnik, K.~S. BKin, C.~Beck, L.~R. Brunham, Diagnostic
  accuracy of ultrasound and mri for achilles tendon xanthoma in people with
  familial hypercholesterolemia: a systematic review, Journal of Clinical
  Lipidology 13~(1) (2019) 40--48.

\bibitem{faust2018comparative}
O.~Faust, U.~R. Acharya, K.~M. Meiburger, F.~Molinari, J.~E. Koh, C.~H. Yeong,
  P.~Kongmebhol, K.~H. Ng, Comparative assessment of texture features for the
  identification of cancer in ultrasound images: a review, Biocybernetics and
  Biomedical Engineering 38~(2) (2018) 275--296.

\bibitem{matthews2018classification}
W.~Matthews, R.~Ellis, J.~Furness, W.~Hing, Classification of tendon matrix
  change using ultrasound imaging: a systematic review and meta-analysis,
  Ultrasound in medicine \& biology 44~(10) (2018) 2059--2080.

\bibitem{tajeddin2018melanoma}
N.~Zamani~Tajeddin, B.~Mohammadzadeh~Asl, Melanoma recognition in dermoscopy
  images using lesion's peripheral region information, Computer Methods and
  Programs in Biomedicine 163 (2018) 143--153.

\bibitem{lecun2015deep}
Y.~LeCun, Y.~Bengio, G.~Hinton, Deep learning, Nature 521~(7553) (2015)
  436--444.

\bibitem{gupta2014curvelet}
R.~Gupta, I.~Elamvazuthi, S.~C. Dass, I.~Faye, P.~Vasant, J.~George, F.~Izza,
  Curvelet based automatic segmentation of supraspinatus tendon from ultrasound
  image: a focused assistive diagnostic method, BioMedical Engineering OnLine
  13~(1) (2014) 157--174.

\bibitem{chang2019quantitative}
R.-F. Chang, C.-C. Lee, C.-M. Lo, Quantitative diagnosis of rotator cuff tears
  based on sonographic pattern recognition, PloS one 14~(2) (2019) e0212741.

\bibitem{meiburger2018quantitative}
K.~Meiburger, M.~Salvi, M.~Giacchino, U.~Acharya, M.~Minetto, C.~Caresio,
  F.~Molinari, Quantitative analysis of patellar tendon abnormality in
  asymptomatic professional “pallapugno” players: A texture-based
  ultrasound approach, Applied Sciences 8~(5) (2018) 660--672.

\bibitem{chuang2014model}
B.-I. Chuang, Y.-N. Sun, T.-H. Yang, F.-C. Su, L.-C. Kuo, I.-M. Jou,
  Model-based tendon segmentation from ultrasound images, in: 40th Annual
  Northeast Bioengineering Conference (NEBEC), IEEE, 2014, pp. 1--2.

\bibitem{cootes2001active}
T.~F. Cootes, G.~J. Edwards, C.~J. Taylor, Active appearance models, IEEE
  Transactions on Pattern Analysis \& Machine Intelligence~(6) (2001) 681--685.

\bibitem{benrabha2017}
J.~Benrabha, F.~Meziane, Automatic roi detection and classification of the
  achilles tendon ultrasound images, in: 1st International Conference on
  Internet of Things and Machine Learning, ACM, 2017, pp. 69--75.

\bibitem{litjens2017survey}
G.~Litjens, T.~Kooi, B.~E. Bejnordi, A.~A.~A. Setio, F.~Ciompi, M.~Ghafoorian,
  J.~A. Van Der~Laak, B.~Van~Ginneken, C.~I. S{\'a}nchez, A survey on deep
  learning in medical image analysis, Medical image analysis 42 (2017) 60--88.

\bibitem{jacobson2011shoulder}
J.~A. Jacobson, Shoulder us: anatomy, technique, and scanning pitfalls,
  Radiology 260~(1) (2011) 6--16.

\bibitem{jahanifar2019supervised}
M.~Jahanifar, N.~Z. Tajeddin, B.~M. Asl, A.~Gooya, Supervised saliency map
  driven segmentation of lesions in dermoscopic images, IEEE Journal of
  Biomedical and Health Informatics 23~(2) (2019) 509--518.

\bibitem{nasnet}
B.~Zoph, V.~Vasudevan, J.~Shlens, Q.~V. Le, Learning transferable architectures
  for scalable image recognition, in: IEEE Conference on Computer Vision and
  Pattern Recognition, 2018, pp. 8697--8710.

\bibitem{xception}
F.~Chollet, Xception: Deep learning with depthwise separable convolutions, in:
  IEEE Conference on Computer Vision and Pattern Recognition, 2017, pp.
  1251--1258.

\bibitem{van2015transfer}
A.~Van~Opbroek, M.~A. Ikram, M.~W. Vernooij, M.~De~Bruijne, Transfer learning
  improves supervised image segmentation across imaging protocols, IEEE
  Transactions on Medical Imaging 34~(5) (2015) 1018--1030.

\bibitem{jahanifar2018segmentation}
M.~Jahanifar, N.~Zamani~Tajeddin, N.~Alemi~Koohbanani, A.~Gooya, N.~Rajpoot,
  Segmentation of skin lesions and their attributes using multi-scale
  convolutional neural networks and domain specific augmentations, arXiv
  preprint arXiv:1809.10243.

\bibitem{deng2009imagenet}
J.~Deng, W.~Dong, R.~Socher, L.-J. Li, K.~Li, L.~Fei-Fei, Imagenet: A
  large-scale hierarchical image database, in: IEEE Conference on Computer
  Vision and Pattern Recognition, IEEE, 2009, pp. 248--255.

\bibitem{uns}
[dataset] Kaggle,
  \href{https://www.kaggle.com/c/ultrasound-nerve-segmentation/data}{Ultrasound
  nerve segmentation}, [Online; accessed 21-September-2018], (2016).
\newline\urlprefix\url{https://www.kaggle.com/c/ultrasound-nerve-segmentation/data}

\bibitem{fcn}
J.~Long, E.~Shelhamer, T.~Darrell, Fully convolutional networks for semantic
  segmentation, in: IEEE Conference on Computer Vision and Pattern Recognition,
  2015, pp. 3431--3440.

\bibitem{unet}
O.~Ronneberger, P.~Fischer, T.~Brox, U-net: Convolutional networks for
  biomedical image segmentation, in: International Conference on Medical Image
  Computing and Computer-Assisted Intervention, Springer, 2015, pp. 234--241.

\bibitem{segnet}
V.~Badrinarayanan, A.~Kendall, R.~Cipolla, Segnet: A deep convolutional
  encoder-decoder architecture for image segmentation, IEEE Transactions on
  Pattern Analysis and Machine Intelligence 39~(12) (2017) 2481--2495.

\bibitem{deeplab}
L.-C. Chen, G.~Papandreou, F.~Schroff, H.~Adam, Rethinking atrous convolution
  for semantic image segmentation, arXiv preprint arXiv:1706.05587.

\bibitem{dumoulin2016guide}
V.~Dumoulin, F.~Visin, A guide to convolution arithmetic for deep learning,
  arXiv preprint arXiv:1603.07285.

\bibitem{kingma2014adam}
D.~P. Kingma, J.~Ba, Adam: A method for stochastic optimization, arXiv preprint
  arXiv:1412.6980.

\bibitem{alemi2019spanet}
N.~Alemi~Koohbanani, M.~Jahanifar, A.~Gooya, N.~Rajpoot, Nuclear instance
  segmentation using a proposal-free spatially aware deep learning framework,
  arXiv preprint arXiv:1908.10356.

\bibitem{resnetv2}
C.~Szegedy, S.~Ioffe, V.~Vanhoucke, A.~A. Alemi, Inception-v4, inception-resnet
  and the impact of residual connections on learning, in: Thirty-First AAAI
  Conference on Artificial Intelligence, 2017.

\bibitem{densenet}
G.~Huang, Z.~Liu, L.~Van Der~Maaten, K.~Q. Weinberger, Densely connected
  convolutional networks, in: IEEE Conference on Computer Vision and Pattern
  Recognition, 2017, pp. 4700--4708.

\bibitem{resnet}
K.~He, X.~Zhang, S.~Ren, J.~Sun, Deep residual learning for image recognition,
  in: IEEE Conference on Computer Vision and Pattern Recognition, 2016, pp.
  770--778.

\bibitem{ROC}
T.~Fawcett, An introduction to roc analysis, Pattern Recognition Letters 27~(8)
  (2006) 861--874.

\bibitem{roy2015diagnostic}
J.-S. Roy, C.~Bra{\"e}n, J.~Leblond, F.~Desmeules, C.~E. Dionne, J.~C.
  MacDermid, N.~J. Bureau, P.~Fr{\'e}mont, Diagnostic accuracy of
  ultrasonography, mri and mr arthrography in the characterisation of rotator
  cuff disorders: a systematic review and meta-analysis, British Journal of
  Sports Medicine 49~(20) (2015) 1316--1328.

\end{thebibliography}

\end{document}